\documentclass[10pt,twocolumn,twoside,journal]{IEEEtran}
\IEEEoverridecommandlockouts
\usepackage{subfigure} 
\usepackage{graphicx}

\usepackage{amsmath,graphicx,amssymb,mathtools,bm}
\usepackage{subfigure}
\usepackage{hyperref}
\usepackage{cite}
\usepackage{amsmath,amssymb,amsfonts}  
\usepackage{textcomp}
\usepackage{xcolor}
\usepackage{verbatim}  
\usepackage{bm}  
\usepackage{mathrsfs} 
\usepackage{algorithmic} 
\usepackage{booktabs}
\usepackage{textcomp}  
\usepackage{multirow}  
\usepackage{lettrine}   
\usepackage{graphicx}  
\usepackage{color}  
\usepackage{amsmath}
\usepackage{amssymb}
\usepackage{stfloats} 
\usepackage{caption} 

\usepackage{color}  
\usepackage[lined,boxed,commentsnumbered]{algorithm2e}

\begin{document}
	\newcommand{\tabincell}[2]{\begin{tabular}{@{}#1@{}}#2\end{tabular}}

 \newtheorem{Proposition}{\bf Proposition}[section]
\newtheorem{Summary}{\bf Summary}
\newtheorem{remark}[Proposition]{Remark}
\newtheorem{Corollary}[Proposition]{Corollary}
\newenvironment{Proof}{{\indent \it Proof:}}{\hfill $\blacksquare$\par}
  
 \title{Flexible Intelligent Metasurface-Aided Wireless Communications: Architecture and Performance}

\author{Songjie Yang, Zihang Wan, Boyu Ning,~\IEEEmembership{Member,~IEEE}, Weidong Mei,~\IEEEmembership{Member,~IEEE},  \\ Jiancheng An,~\IEEEmembership{Member,~IEEE}, Yonina C. Eldar,~\IEEEmembership{Fellow,~IEEE}, and 
	Chau Yuen,~\IEEEmembership{Fellow,~IEEE}

\thanks{
	Songjie Yang,   Boyu Ning, and Weidong Mei  are with the National Key Laboratory of Wireless Communications, University of Electronic Science and Technology of China, Chengdu 611731, China. (e-mail:	yangsongjie@std.uestc.edu.cn; boydning@outlook.com; wmei@uestc.edu.cn). 

Zihang Wan, is with the Key Laboratory of Wireless Optical Communication, Chinese Academy of Sciences, School of Information Science and Technology, University of Science and Technology of China, Hefei 230026, China (e-mail: zihangwan@mail.ustc.edu.cn)

Yonina C. Eldar is with the Faculty of Mathematics and Computer
Science, Weizmann Institute of Science, Rehovot 7610001, Israel (e-mail:
yonina.eldar@weizmann.ac.il).

Jiancheng An and Chau Yuen are with the School of Electrical and Electronics Engineering, Nanyang Technological University, 639798, Singapore (e-mail: jiancheng an@163.com; chau.yuen@ntu.edu.sg).
	}
}
\maketitle

\begin{abstract}
Typical reconfigurable intelligent surface (RIS) implementations include
metasurfaces with almost passive unit
elements capable of reflecting their
incident waves in controllable ways, enhancing wireless communications in a cost-effective manner.
 In this paper, we advance the concept of intelligent {metasurfaces} by introducing a flexible array geometry, termed  flexible intelligent metasurface (FIM), which supports both element movement (EM) and passive beamforming (PBF). In particular, based on the single-input single-output (SISO) system setup, we first compare three modes of FIM, namely, EM-only, PBF-only, and EM-PBF, in terms of received signal power under different FIM and channel setups. {The} PBF-only mode, which only adjusts the reflect phase, is shown to be less effective than the EM-only mode in enhancing received signal strength. In a multi-element, multi-path scenario, the EM-only mode improves the received signal power by $125\%$ compared to the PBF-only mode. The EM-PBF mode, which optimizes both element positions and phases, further enhances performance. Additionally, we investigate the channel estimation problem for FIM systems by designing a protocol that gathers EM and PBF measurements, enabling the formulation of a compressive sensing problem for joint cascaded and direct channel estimation. {We then propose} a sparse recovery algorithm called clustering mean-field variational sparse Bayesian learning, {which enhances estimation performance while maintaining low complexity.}
\end{abstract}
\section{Introduction}  
As wireless communication technologies progress, there is a growing emphasis on achieving greater degrees of freedom (DoFs) and expanding its applications.  A key area of focus is the exploration of wireless channel characteristics to enhance both communication and sensing capabilities.  This includes leveraging the sparsity of millimeter-wave and Terahertz channels for beamspace signal processing \cite{mmw1,mmw2,MEI2}, using reconfigurable intelligent surfaces (RISs) to improve channel propagation \cite{RIS1,RIS3,MEI1}, and exploiting near-field spherical-wave channels to enhance distance DoF \cite{NF1,NF2}. Despite significant advancements, the potential of wireless channels remains vast, offering promising opportunities for further research and innovation.
 
RISs, which possess the ability to reflect, refract, and manipulate incoming electromagnetic waves \cite{RIS1,RIS3,MEI1}, have emerged as a promising solution to address the challenges of wireless propagation. Their capabilities have opened new avenues for research in areas such as channel estimation, beamforming, and localization. While single-RIS systems have been widely explored, significant attention has also been given to double- and multi-RIS configurations, where multi-reflection among RISs can more effectively create blockage-free line-of-sight (LoS) links in complex wireless environments \cite{double-RIS1,double-RIS2,multi-RIS1}. For instance, one method involves stacking multiple metasurfaces into a centralized system, termed stacked intelligent metasurfaces, to achieve efficient analog signal processing in the wave domain \cite{CRIS1}. Another strategy involves the deployment of multi-sector RISs to achieve  the beyond diagonal phase shift matrix for enhancing the coverage \cite{CRIS2}. {The} potential of RIS technology is further amplified by advances in electromagnetic information theory, exemplified by concepts such as holographic RIS \cite{HOL1,HOL2} and near-field RIS \cite{XL-RIS1,XL-RIS2}.

In addition to RIS, directly changing the transceiver's array geometry can also significantly affect the channel, introducing greater flexibility to antenna arrays. This concept originating from antenna selection \cite{AS} aims to find a given number of optimal antenna positions within a candidate set based on specific selection criteria. The antenna selection process can be regarded as a discrete optimization problem.
A related concept in the field of array signal processing is array synthesis. This approach aims to optimize antenna excitation coefficients---similar to beamforming/precoding---as well as array structures, including antenna positions and orientations, to achieve specific radiation patterns such as enhancing beam directivity, suppressing sidelobes, and creating nulls \cite{SAS1,SAS2}. 
Antenna selection is usually achieved through electronic switches, inherent properties of antennas. Although array synthesis involves physical movement of the antennas or equivalent electronic phase control \cite{Stepper}, it remains largely within the domain of antenna and microwave engineering and has not yet garnered significant attention in wireless communications.

The recent decades have witnessed the development of shape-adjustable, controllable, and reconfigurable materials and antennas. These advancements, largely driven by researchers in the antenna domain, are finding applications across various fields, including wearable devices and dynamic deployment designs. Progress in flexible antennas, encompassing position movement, shape control, rotation adjustment, and pattern reconfiguration, has been well studied in \cite{Liquid, MEMS, Pixel}.
 In this context, the concept of flexible antennas has attracted significant attention in wireless communications. Recent studies have shown the potential of leveraging the mobility of antennas within a confined area for enhancing wireless communication performance. For example, fluid antenna systems (FASs) and movable antennas (MAs) have received considerable interest for their role in elucidating how element movement affects wireless communications in various channel conditions \cite{FAS1,FAS2,MA0,MAM,Z1,Z2}. Several algorithms have been proposed to optimize the antenna positions to characterize the theoretical performance limits of FASs/MAs, including convex optimization \cite{MA1}, gradient-based search \cite{MA2,MA3}, graph-based algorithms \cite{MA4}, and sparse optimization \cite{MA5}. Considering practical implementations, the authors in \cite{MAM} proposed several general MA architectures beyond element movement to cater to specific application. Moreover, the authors in \cite{F2} examined the ability of flexible arrays to rotate, bend, and fold for maximizing multi-user sum-rate in the cellular netowork.

Given the potential of RISs and flexible arrays in configuring channel conditions, a straightforward method is to combine them to reap their complementary performance gain. For example, the authors in \cite{MA-RIS1, MA-RIS2, MA-RIS3} {investigated} RIS-aided MAs/FASs, where the RIS {is} utilized to create a virtual LoS channel condition, while the MAs/FASs equipped at the base station (BS) or at the user side {is} utilized for further channel reconfiguration. Inspired by FASs and MAs, several new RIS architectures have also been proposed in the literature, e.g., dynamically rotatable RIS \cite{RIS2} and element-position-movable RIS \cite{MA-RIS4, MA-RIS5}, which are able to adjust their element positions and perform passive beamforming at the same time. Despite these advancements, the integration of element mobility into RISs is still in its infancy and requires more in-depth study to {realize} its full potential.

{Against this background, this paper explores the use of flexible intelligent metasurfaces (FIM) with both passive beamforming (PBF) and element movement (EM) techniques. The study focuses on: 1) determining whether these methods can synergize to achieve results greater than the sum of their individual contributions, 2) assessing whether they can substitute for each other given their similar functions, and 3) understanding how to acquire channel state information for these types of reconfigurable intelligent surfaces (RIS).}
 The main contributions are as follows\footnote{The source code for this work is openly available at \url{https://github.com/YyangSJ/Flexible-RIS}}.

 \begin{itemize}
 \item 
{We investigate a point-to-point FIM-aided single-input single-output (SISO) setup, evaluating the performance of the FIM across three distinct modes: EM-only, PBF-only, and a combined EM-PBF mode. Our focus is on maximizing the received signal power for each of these modes under both single-path and multi-path channel conditions, and with various FIM configurations. By adjusting the positions of the FIM elements and their phase coefficients, we demonstrate the effects of spatial interference. For simple scenarios, such as single-path channels, we derive analytical solutions for optimizing EM and PBF. For more complex cases, we employ the Bayesian optimization method to fine-tune the FIM element positioning and phase adjustments. We compare and analyze the three modes against the derived upper bound of the received power, offering insights into their relative performance.}
 
 \item   
{We propose a channel estimation protocol for FIM where $Q$ EM actions are divided into $Q$ subframes, each  containing $T$ time slots for PBF, resulting in a $QT$-dimensional spatial measurement for channel estimation. Our proposed estimation framework simultaneously estimates direct and cascaded channels. 
 Building on this framework, we introduce the clustering mean-field variational sparse Bayesian learning (CMFV-SBL) algorithm. This algorithm enhances previous V-SBL and MFV-SBL approaches by incorporating a partially factorized form using clustering methods. Unlike the fully factorized MFV-SBL, which ignores inter-atom relationships, CMFV-SBL employs K-means clustering to group atoms into clusters, thereby constructing a clustering factorized form over variational distributions. This method efficiently captures atom relationships, especially with a correlated sensing matrix.}

 \end{itemize}

 {\emph {Notations}}:  ${\left(  \cdot  \right)}^{ *}$, ${\left(  \cdot  \right)}^{T}$,
 ${\left(  \cdot  \right)}^{ H}$, and $\left(\cdot\right)^{-1}$ denote  conjugate, transpose, conjugate transpose, and inverse, respectively. $\vert\cdot\vert$ represents the modulus. $\Vert\mathbf{a}\Vert_2$ and
 $\Vert\mathbf{A}\Vert_F$ denotes the $\l_2$ norm of vector $\mathbf{a}$ and Frobenius norm of matrix $\mathbf{A}$,   $\rm Tr(\cdot)$ denotes the trace. $\circ$ represents the Hadamard product, $\Re\{\cdot\}$ denotes the real part of a complex-value number, and
 $[\mathbf{A}]_{i,:}$ and $[\mathbf{A}]_{:,j}$ denote the $i$-th row and the $j$-th column of matrix $\mathbf{A}$, respectively. Moreover, $\mathbb{E}\{\cdot\}$ is the expectation,  $\mathbf{I}_K$ is a $K\times K$ identity matrix, and $\mathcal{CN}(\mathbf{a},\mathbf{A})$ is the complex Gaussian distribution with mean $\mathbf{a}$ and covariance matrix $\mathbf{A}$.
\section{System Model}
  
\begin{figure}
	\centering 
	\includegraphics[width=3.3in]{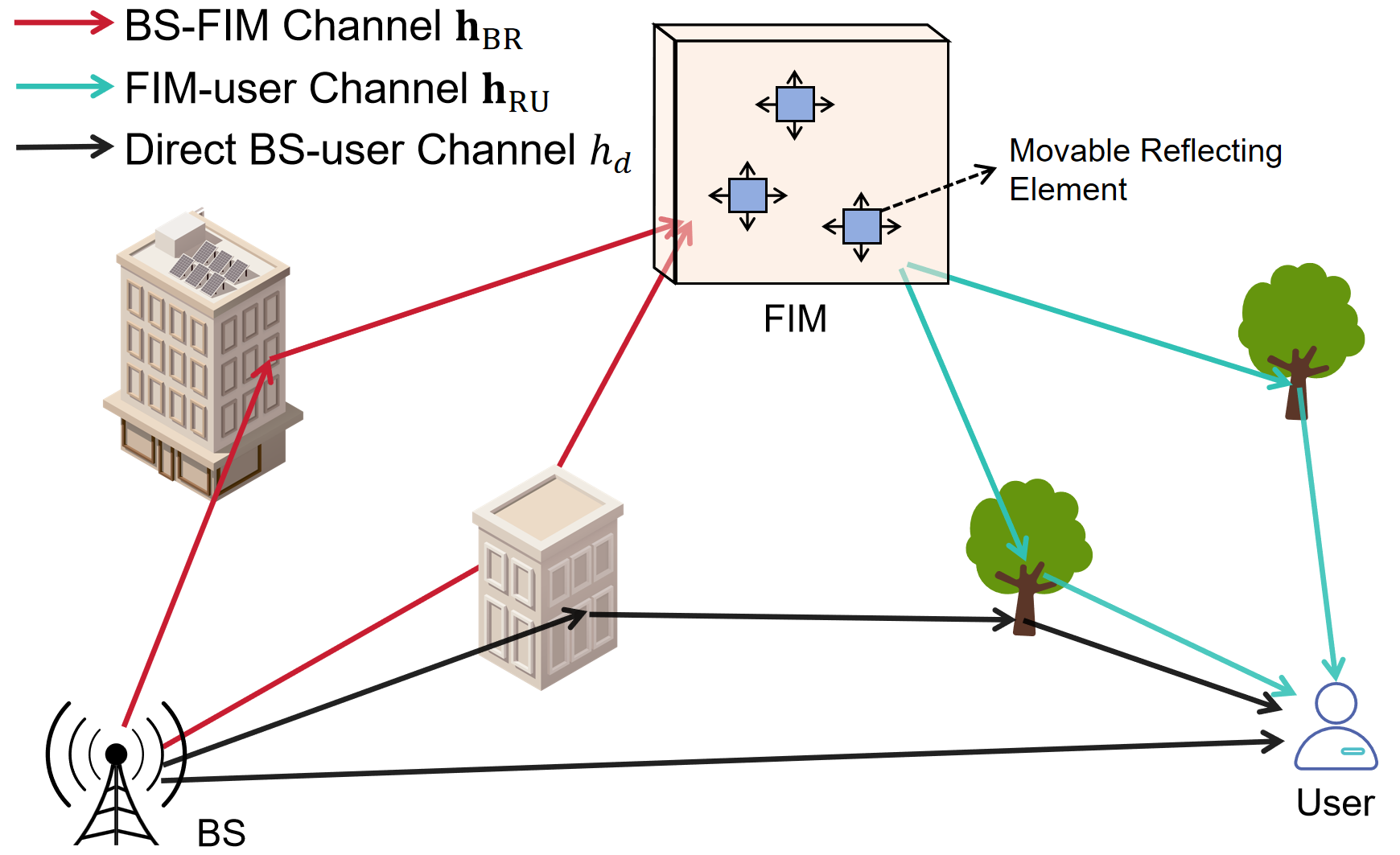}
	\caption{Schematic of a FIM aided communication.}\label{sys} 
\end{figure}

{In this work, we consider a point-to-point FIM-aided SISO communication system, as shown in Fig. \ref{sys}. The FIM is equipped with $N$ elements capable of moving within the $x$-$z$ plane and reflecting signals.  $\{x_n,z_n\}_{n=1}^N$ represent positions for the $N$ elements.  Considering the feasible position constraints for EM, we examine the square area $\mathcal{R}=\left\{(x,z) \mid x,z \in [-R,R]\right\}$, where $R$ denotes the boundary along the $x$- and $z$-axes. 
	 We employ PBF in this system, which is achieved through phase adjustment without additional energy sources. 
	 The PBF matrix is defined as $\mathbf{V}={\rm diag}\{e^{jv_1},e^{jv_2},\cdots,e^{jv_N}\}\in\mathbb{C}^{N\times N}$, with $\mathbf{v}=[v_1,v_2,\cdots,v_N]^T$ being the phase vector of the FIM.  The FIM is allowed to operate in EM-only, PBF-only, or EM-PBF modes in different scenarios, optimizing for either energy efficiency or performance.}

{As the BS transmits signals, the user receives these signals through a multipath channel, described by $h\triangleq \mathbf{h}_{\rm FU}^H\mathbf{V}\mathbf{h}_{\rm BF} + h_{d}$. As shown in Fig. \ref{sys}, the channel between the BS and the FIM, the channel between the FIM and the user, and the direct channel between the BS and the user are represented by $\mathbf{h}_{\rm BF}$, $\mathbf{h}_{\rm FU}$, and ${h}_{d}$, respectively. Table \ref{table1} lists  symbols used through this paper.}

	Considering the multipath channel model, we assume that the channel $\mathbf{h}_{\rm BF}$ consists of $L$ paths and the channel $\mathbf{h}_{\rm FU}$ consists of $P$ paths. These channels can be defined as follows:
	\begin{equation}\label{hb}
		\mathbf{h}_{\rm BF}=\sqrt{\frac{N}{L}} \sum_{l=1}^{L}\alpha_l \mathbf{a}_N(\theta_{{\rm B},l},\phi_{{\rm B},l}),
	\end{equation}
	where $\{\alpha_l\}_{l=1}^L$ denotes the complex path gains,  $\theta_{{\rm B},l}\triangleq \cos(\vartheta_{{\rm B},l})$ and $\phi_{{\rm B},l}\triangleq \sin(\vartheta_{{\rm B},l})\cos(\varphi_{{\rm B},l})$, $\forall l$, are the virtual angles, respectively. $\vartheta_{{\rm B},l}$ and $\varphi_{{\rm B},l}$ are the physical elevation and azimuth angles of the $l$-th path of the BS-FIM channel, respectively.  The array manifold $\mathbf{a}$ is assumed to follow the far-field planar wavefront:
	\begin{equation}
		\mathbf{a}_N(\theta,\phi)=\sqrt{\frac{1}{N}}\left[e^{j\frac{2\pi}{\lambda}(x_1\phi+z_1\theta)},\cdots, e^{j\frac{2\pi}{\lambda}(x_N\phi+z_N\theta)}\right]^T,
	\end{equation}
	where $\mathbf{x}\triangleq[x_1,\cdots,x_N]^T\in\mathbb{C}^{N\times 1}$ and $\mathbf{z}\triangleq[z_1,\cdots,z_N]^T\in\mathbb{C}^{N\times 1}$ represent the $N$ elements' coordinates in the $x$-$z$ plane, respectively.
	
	Similarly, the FIM-user channel $\mathbf{h}_{\rm FU}$ is expressed by
	\begin{equation}\label{hu}
		\mathbf{h}_{\rm FU}= \sqrt{\frac{N}{P}}\sum_{p=1}^{P}\beta_p \mathbf{a}_N(\theta_{{\rm U},p},\phi_{{\rm U},p}),
	\end{equation}
	where $\{\beta_p\}_{p=1}^P$ denotes the complex path gains, $\theta_{{\rm U},l}\triangleq \cos(\vartheta_{{\rm U},l})$ and $\phi_{{\rm U},l}\triangleq \sin(\vartheta_{{\rm U},l})\cos(\varphi_{{\rm U},l})$, $\forall l$, are the virtual angles, respectively, and $\vartheta_{{\rm U},l}$ and $\varphi_{{\rm U},l}$ are the physical elevation and azimuth angles of the $l$-th path of the FIM-user channel, respectively.
	
	\begin{table}[]
		\caption{Variable notation.}
		\begin{tabular}{|l|l|}
			\hline
			$N$                                                                                                                   & Number of FIM elements                                                                                                                                                                       \\ \hline
			$\lambda$                                                                                                             & Antenna wavelength                                                                                                                                                                           \\ \hline
			$\mathbf{x}$                                                                                                          & FIM's element position along $x$-axis                                                                                                                                                        \\ \hline
			$\mathbf{z}$                                                                                                          & FIM's element position along $z$-axis                                                                                                                                                        \\ \hline
			$\mathbf{v}$                                                                                                          & FIM's element phase                                                                                                                                                                          \\ \hline
			$\mathbf{h}_{\rm BF}$                                                                                                 & BS-FIM channel                                                                                                                                                                               \\ \hline
			\begin{tabular}[c]{@{}l@{}}$L,\alpha,\vartheta_{\rm B},$\\ $\varphi_{\rm B},\theta_{\rm B},\phi_{\rm B}$\end{tabular} & \begin{tabular}[c]{@{}l@{}}Number of paths, path gain, physical elevation angle,\\ physical azimuth angle, virtual elevation, and virtual\\ azimuth angle of the BS-FIM channel\end{tabular} \\ \hline
			$\mathbf{h}_{\rm FU}$                                                                                                 & FIM-user channel                                                                                                                                                                             \\ \hline
			\begin{tabular}[c]{@{}l@{}}$P,\beta,\vartheta_{\rm U},$\\ $\varphi_{\rm U},\theta_{\rm U},\phi_{\rm U}$\end{tabular}  & \begin{tabular}[c]{@{}l@{}}Number of paths, path gain, physical elevation angle,\\ physical azimuth angle, virtual elevation, and virtual\\ azimuth angle of the BS-FIM channel\end{tabular} \\ \hline
			$h_d$                                                                                                                 & Direct BS-user channel                                                                                                                                                                       \\ \hline
			$\gamma$                                                                                                              & Path gain of the BS-user channel                                                                                                                                                             \\ \hline
			$h_{\rm cas}$                                                                                                         & Cascaded channel                                                                                                                                                                             \\ \hline
			$\widetilde{\theta}$, $\widetilde{\phi}$                                                                              & Cascaded elevation and azimuth angles                                                                                                                                                        \\ \hline
			$\mathbf{a}$                                                                                                          & Far-field array manifold                                                                                                                                                                     \\ \hline
			$s$                                                                                                                   & Transmitted signal                                                                                                                                                                           \\ \hline
			$n$                                                                                                                   & Noise                                                                                                                                                                                        \\ \hline
		\end{tabular}
		\label{table1}
	\end{table}
	Given $\mathbf{h}_{\rm FR}$ and $\mathbf{h}_{\rm FU}$, the cascaded channel $h_{\rm cas}$ is given by  
	\begin{equation}
		\begin{aligned}
			h_{\rm cas}&= \mathbf{h}_{\rm FU}^H\mathbf{V}\mathbf{h}_{\rm BF} \\
			&=  \mathbf{v}^T {\rm diag}\left(\mathbf{h}_{\rm FU}^H\right)\mathbf{h}_{\rm BF} \\
			&=\frac{N}{\sqrt{LP}} \mathbf{v}^T \left( \left( \sum_{p=1}^P \beta_p^* \mathbf{a}^*_N(\theta_{{\rm U},p},\phi_{{\rm U},p})  
			\right) \right. \\  &\ \ \ \ \ \ \ \ \ \ \ \ \ \ \ \ \ \  \left. 
			\circ\left(\sum_{l=1}^{L}\alpha_l \mathbf{a}_N(\theta_{{\rm B},l},\phi_{{\rm B},l})\right)\right)
			\\
			& = \sqrt{\frac{N}{LP}}\sum_{l=1}^L\sum_{p=1}^P \alpha_l\beta^*_p 
			\mathbf{v}^T \mathbf{a}_N(\theta_{{\rm B},l}-\theta_{{\rm U},p},\phi_{{\rm B},l}-\phi_{{\rm U},p})\\
			& =\sqrt{\frac{N}{LP}}\sum_{l=1}^L\sum_{p=1}^P \alpha_l\beta^*_p 
			\mathbf{v}^T \mathbf{a}_N\left(\widetilde{\theta}_{l,p},\widetilde{\phi}_{l,p}\right),
		\end{aligned}
	\end{equation}  
	where $\widetilde{\theta}_{l,p}\triangleq \theta_{{\rm B},l}-\theta_{{\rm U},p}$ and $\widetilde{\phi}_{l,p}\triangleq\phi_{{\rm B},l}-\phi_{{\rm U},p}$ are defined.
	
	Moreover, the direct channel $h_d\triangleq\gamma$ is assumed to follow a complex Gaussian distribution.

 Then, the received signal $y$ at the user is expressed as
\begin{equation}\label{ye}
	y=hs+n,
\end{equation}
where $s$ is the transmitted signal with the assumption of $\mathbb{E}\{\vert s\vert^2\}=1$, and $n$ is noise distributed as $\mathcal{CN}(0,\sigma^2_n)$.

{The following section evaluates the received power $\mathbb{E}\{|y|^2\}$ based on  (\ref{ye}) by optimizing $\{x_n,z_n,v_n\}_{n=1}^N$. This expression is simplified into $|h|^2+\sigma_n^2$, given that the noise and the transmitted signal are independently distributed.}

\section{EM and PBF Analysis}
This section investigates the impact of EM and PBF on the received power, considering three different modes: EM-only, PBF-only, and EM-PBF, in various scenarios.

\subsection{Single-Element Single-Path}
{In the single-element single-path case where $N=1$, we ignore the subscript $n$, and focus on optimizing parameters  $\{v,z,x\}$ to maximize the received power. First,
 the cascaded channel is expressed as
$h_{\rm cas}=\alpha\beta^* e^{jv}e^{j\frac{2\pi}{\lambda}\left( \widetilde{\theta}z+\widetilde{\phi}x\right)}$. Next, the problem of maximizing the received power with respect to $\{v,z,x\}$ is formulated as follows:  }
\begin{equation}\label{fx1}
	\begin{aligned}
		\underset{v,z,x}{{\rm  arg\ max}}  \ f(v,x,z)&\triangleq \vert h_{\rm cas}+h_d \vert ^2  \\
		&= \left\vert\alpha\beta^* e^{jv}e^{j\frac{2\pi}{\lambda}\left(\widetilde{\theta}z+\widetilde{\phi}x\right)}+\gamma \right\vert^2.
	\end{aligned}
\end{equation}

\begin{Proposition}\label{Pro3}
{	$\{v,z,x\}$ in problem (\ref{fx1}) can be solved by addressing the following equation:}
\begin{equation}\label{E1}
	v +\frac{2\pi}{\lambda} \left(\widetilde{\theta}z+\widetilde{\phi}x\right) =	\angle \gamma-\angle \alpha \beta^*+2k\pi, \ k\in \mathbb{Z},
\end{equation}
and the objective's upper bound is $	f(v,x,z)\leq\left( |\alpha\beta^*| + |\gamma|\right)^2$.
\end{Proposition}
\begin{Proof}
{Note that, in problem (\ref{fx1}), the maximum value is achieved when  $h_{\rm cas}$ and $h_d$ are phase-aligned, such that $\angle h_{\rm cas}=\angle h_d$. In this sense, the optimal objective $\left( |\alpha\beta^*| + |\gamma|\right)^2$ is reached, resulting in constructive interference. Conversely, when destructive interference occurs, characterized by $\angle h_{\rm cas}=-\angle h_d$, $f(v,x,z)$ reaches its minimum value of $\left( \vert\alpha\beta^*\vert-\vert\gamma\vert \right)^2$.}
\end{Proof}

\begin{Corollary}\label{MR0}
For traditional RIS systems with   fixed element position, i.e., assuming $x=z=0$, the optimal reflective phase is obtained by $v=\angle \gamma-\angle \alpha \beta^*+2k\pi$.  
\end{Corollary}

\begin{Corollary}\label{MR1}
	For the EM-only case, by setting $v=0$, the maximum path gain is achieved when $\{x,z\}$ satisfy $\frac{2\pi}{\lambda} \left(\widetilde{\theta}z+\widetilde{\phi}x\right) = \angle \gamma - \angle \alpha \beta^* + 2k\pi$, following a periodic line equation. The minimum period is $\frac{\lambda}{\sqrt{\widetilde{\theta}^2+\widetilde{\phi}^2}}$, which is the distance between adjacent lines.
\end{Corollary}
 
According to Corollaries \ref{MR0} and \ref{MR1}, in FIM systems with a single-element single-path case, both adjusting the phase coefficient and element position can achieve optimal performance. Thus, EM and PBF can substitute for each other, and their collaboration does not yield additional gains.

\begin{figure}
 	\centering 
 	\includegraphics[width=3.7in]{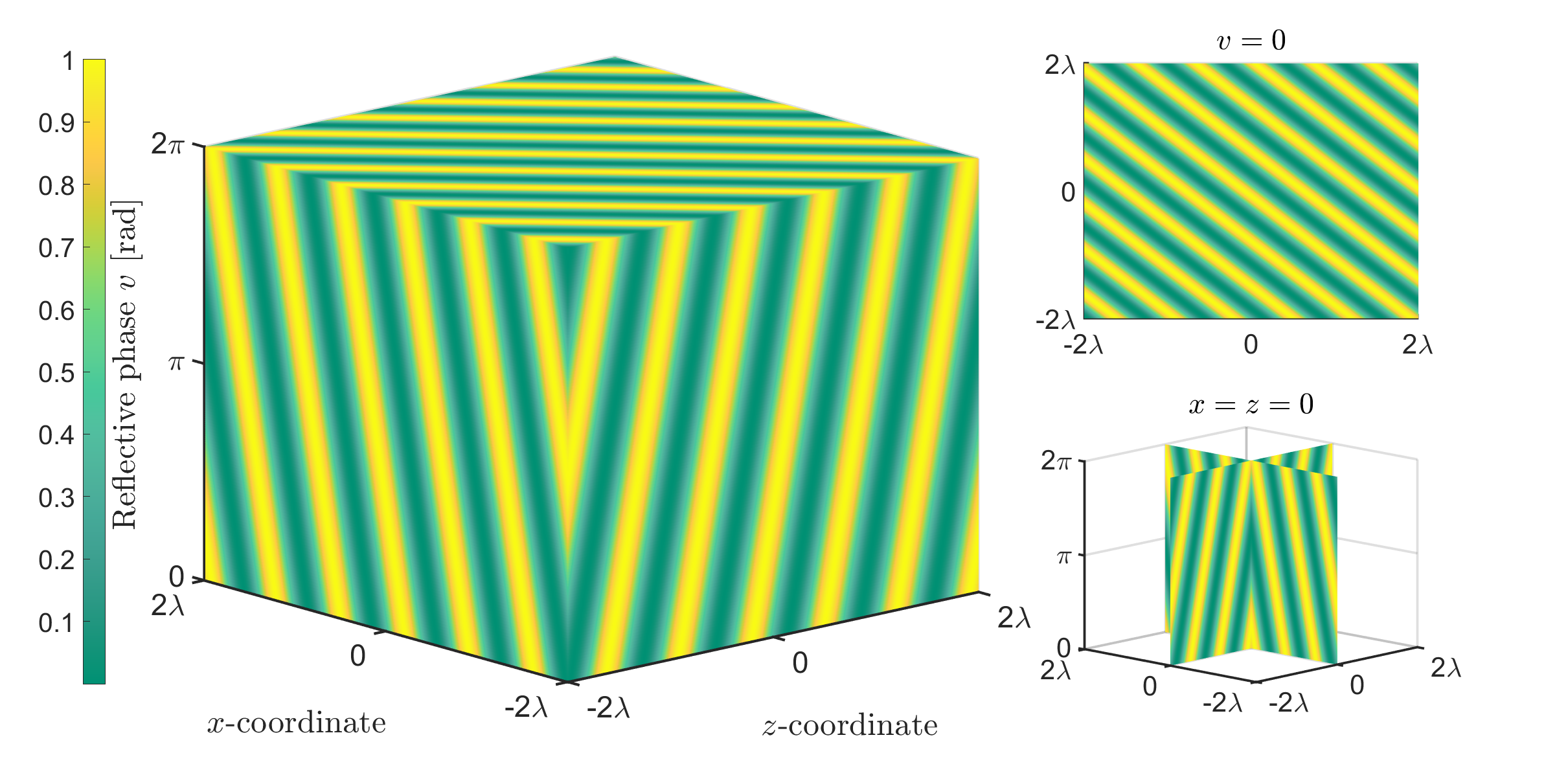}
 	\caption{{Spatial} constructive/destructive interference with respect to $\{v,x,z\}$ in the single-element single-path case.}\label{S-SP} 
 \end{figure}

The interference effect described in Proposition \ref{Pro3} is visualized in Fig. \ref{S-SP}, where the parameters are set as follows: ${\theta}_{\rm B}=\frac{\sqrt{2}}{2}$, ${\theta}_{\rm U}=-\frac{\sqrt{2}}{2}$, ${\phi}_{\rm B}=\frac{\sqrt{2}}{2}$, ${\phi}_{\rm U}=-\frac{\sqrt{2}}{2}$, ${\alpha}=e^{j\frac{\pi}{4}}$, ${\beta}=e^{j\frac{\pi}{4}}$, ${\gamma}=e^{j\frac{\pi}{4}}$, $\lambda=0.03$ meters, and the boundary $R=2\lambda$. The left side of Fig. \ref{S-SP} illustrates the spatial constructive and destructive interference fringes in three dimensions, with yellow and green fringes representing $\{v,x,z\}$ values that satisfy the normalized maximum and minimum objectives, respectively. In this parameter setting, the constructive interference forms an inverted triangle. Additionally, the top-right and bottom-right sections of Fig. \ref{S-SP} show how the objective function value varies with $\{x,z\}$ (when $v=0$) and with $v$ (when $x=z=0$), respectively. The optimal values for both the EM-only and PBF-only cases can be identified from these plots.
 
 \subsection{Multi-Element Single-Path}
 The cascaded channel in the  multi-element single-path case is described by
$
 	h_{\rm cas}=\alpha\beta^*\sum_{n=1}^{N} e^{jv_n}e^{j\frac{2\pi}{\lambda}\left(\widetilde{\theta}z_n+\widetilde{\phi}x_n\right)} 
$, yielding the objective:
 \begin{equation}\label{hcdms}
 	\begin{aligned}
 		f(v,\mathbf{x},\mathbf{z}) =\left\vert\alpha\beta^*\sum_{n=1}^{N} e^{jv_n}e^{j\frac{2\pi}{\lambda}\left(\widetilde{\theta}z_n+\widetilde{\phi}x_n\right)} +\gamma \right\vert^2.  
 	\end{aligned}
 \end{equation}
 
 \begin{Proposition}\label{P34}
{Maximizing (\ref{hcdms}) yields the following equation system to optimize $\{v,\mathbf{x},\mathbf{z}\}$:}
 	\begin{equation}\label{esy1}
 		\begin{cases}
 			v_1 +\frac{2\pi}{\lambda} \left(\widetilde{\theta}z_1+\widetilde{\phi}x_1\right)=\angle\gamma-\angle \alpha \beta^* +2k_1\pi, \\ v_2 +\frac{2\pi}{\lambda} \left(\widetilde{\theta}z_2+\widetilde{\phi}x_2\right)= \angle \gamma-\angle \alpha \beta^*+2k_2\pi,
 			\\ \ \ \ \ \ \ \ \ \ \ \ \ \ \ \ \ \ \ \ \ \ \ \ \ \ \ \ \vdots \\ v_N +\frac{2\pi}{\lambda} \left(\widetilde{\theta}z_N+\widetilde{\phi}x_N\right)= \angle \gamma-\angle \alpha \beta^*+2k_N\pi,
 		\end{cases}
 	\end{equation} 
 where $k_n\in\mathbb{Z}, \forall n\in\{1,\cdots,N\}$. Moreover, the objective's upper bound is $	f(v,\mathbf{x},\mathbf{z})\leq  \left( N|\alpha\beta^*| + |\gamma|\right)^2$.
 \end{Proposition}
 
 \begin{Proof}
 	By aligning the phase $\angle h_{\rm cas}=\angle h_d$, we have $\angle\left( \alpha\beta^\star e^{jv_n}e^{j\frac{2\pi}{\lambda}\left(\widetilde{\theta}z_n+\widetilde{\phi}x_n\right)}\right)=\angle \gamma$, for $\forall n$, resulting in (\ref{esy1}). Meanwhile, the objective's maximum is reached by factoring out the phase. 
 \end{Proof}
\begin{Corollary}
	For traditional RIS systems with fixed antenna positions, the reflective phase can be obtained by $ v_n= \angle \gamma-\angle \alpha \beta^*+2k_n\pi-\frac{2\pi}{\lambda} \left(\widetilde{\theta}\overline{z}_n+\widetilde{\phi}\overline{x}_n\right)
	$, where $\overline{x}_n$ and $\overline{z}_n$ are the fixed element positions following a half-wavelength spacing in a uniform planar array.
\end{Corollary}
\begin{Corollary}
	Similar to \emph{Corollary \ref{MR1}}, the reflective phase can be set to $0$ so that only optimizing $\{x,z\}$ to attain the maximal gain, i.e., satisfying
	\begin{equation}
	\frac{2\pi}{\lambda} \left(\widetilde{\theta}z_n+\widetilde{\phi}x_n\right)=\angle\gamma-\angle \alpha \beta^* +2k_n\pi, \ \forall n.
	\end{equation}
	 The simplest case is $k_n=0,\forall n$, such that $\{x_n,z_n\}_{n=1}^N$ are distributed in a line in intercept form: 
	 \begin{equation}
	 	\frac{z}{a}+ 	\frac{x}{b}=1,
	 \end{equation}
 where $a\triangleq\frac{\lambda(\angle\gamma-\angle \alpha \beta^*)}{2\pi \widetilde{\theta}}$ and $b\triangleq\frac{\lambda(\angle\gamma-\angle \alpha \beta^*)}{2\pi \widetilde{\phi}}$ denote the $x$- and $y$-intercept, respectively. 
\end{Corollary}

Both EM-only and PBF-only modes can achieve optimal performance in the multi-element single-path case, provided that the movable region is sufficiently large to accommodate the minimum inter-element spacing for all elements.

\subsection{Single-Element Multi-Path}
In the single-element single-path case, the cascaded channel is expressed as
$	h_{\rm cas}= e^{jv}\sum_{l=1}^{L}\sum_{p=1}^{P} \alpha_l\beta^*_p e^{j\frac{2\pi}{\lambda}\left(\widetilde{\theta}_{l,p}z+\widetilde{\phi}_{l,p}x\right)}$, yielding the objective:
\begin{equation}\label{hcd2}
	\begin{aligned}
		f(v,x,z) = \left\vert \frac{1}{\sqrt{LP}} e^{jv}\sum_{l=1}^{L}\sum_{p=1}^{P} \alpha_l\beta^*_p e^{j\frac{2\pi}{\lambda}\left(\widetilde{\theta}_{l,p}z+\widetilde{\phi}_{l,p}x\right)}+\gamma \right\vert^2  
	\end{aligned}
\end{equation} 
 
\begin{Proposition}\label{MS_P}
{Maximizing (\ref{hcd2}) yields the following equation system to optimize $\{v, {x}, {z}\}$:}
	\begin{equation}\label{MS}
		\begin{cases}
			v +\frac{2\pi}{\lambda} \left(\widetilde{\theta}_{1,1}z+\widetilde{\phi}_{1,1}x\right)=\angle\gamma-\angle \alpha_1 \beta^*_1 +2k_1\pi, \\ v +\frac{2\pi}{\lambda} \left(\widetilde{\theta}_{1,2}z+\widetilde{\phi}_{1,2}x\right)= \angle \gamma-\angle \alpha_1 \beta^*_2+2k_2\pi,
			\\ \ \ \ \ \ \ \ \ \ \ \ \ \ \ \ \ \ \ \ \ \ \ \ \ \ \ \ \ \ \ \vdots \\ v +\frac{2\pi}{\lambda} \left(\widetilde{\theta}_{L,P}z+\widetilde{\phi}_{L,P}x\right)= \angle \gamma-\angle \alpha_L \beta^*_P+2k_{LP}\pi, 
		\end{cases}
	\end{equation} 
where $\{k_1,\cdots,k_{LP}\}\in\mathbb{Z}$. Moreover, the objective's upper bound is $ f(v,x,z)\leq\left( \frac{1}{\sqrt{LP}} \sum_{l=1}^{L}\sum_{p=1}^{P}|\alpha_l\beta^*_p| + |\gamma|\right)^2$.

\end{Proposition}

\begin{Proof}
This proof closely resembles the proof of Proposition \ref{P34}, employing a similar phase alignment strategy.
\end{Proof}

\begin{Corollary}
	For traditional RIS systems with fixed element positions, (\ref{hcd2}) simplifies into
	\begin{equation}\label{FBF}
	 f_{\rm PBF}(v) \triangleq \left\vert \frac{1}{\sqrt{LP}} e^{jv}\sum_{l=1}^{L}\sum_{p=1}^{P} \alpha_l\beta^*_p +\gamma\right\vert^2,
	\end{equation}
and its solution is obtained by 
	
	\begin{equation}\label{vs}
		\begin{aligned}
			v  
			=  \angle \gamma-\angle\left( \sum_{l=1}^{L}\sum_{p=1}^{P} \alpha_l\beta^*_p\right).
		\end{aligned}
	\end{equation}
\end{Corollary}

\begin{Corollary}
	For EM-only FIM, (\ref{hcd2}) simplifies into
	\begin{equation}\label{FEM}
		f_{\rm EM}(x,z) \triangleq \left\vert  \frac{1}{\sqrt{LP}} \sum_{l=1}^{L}\sum_{p=1}^{P} \alpha_l\beta^*_p e^{j\frac{2\pi}{\lambda}\left(\widetilde{\theta}_{l,p}z+\widetilde{\phi}_{l,p}x\right)}+\gamma \right\vert^2.
	\end{equation}

	In the case of $LP=2$, and $L=1$ and $P=2$ are set without losing generality, $\{x_k,z_k\}$ without region constraints can be solved by a square system:
	\begin{equation}\label{sqs}
		\begin{cases}
		\frac{2\pi}{\lambda} \left(\widetilde{\theta}_{1,1}z+\widetilde{\phi}_{1,1}x\right)=\angle \gamma-\angle \alpha_1 \beta^*_1 +2k_1\pi, \\ \frac{2\pi}{\lambda} \left(\widetilde{\theta}_{1,2}z+\widetilde{\phi}_{1,2}x\right)= \angle \gamma-\angle \alpha_1 \beta^*_2+2k_2\pi,
		\end{cases}
	\end{equation} 
where $\{x,z\}$ can be accurately solved for $\forall k_1,k_2$, and $k_1$ and $k_2$ can be adjusted for satisfying the allowed movable region.
	
For $LP > 2$,  (\ref{MS}) does not have a unique solution. While the least squares method may provide a feasible solution, it cannot guarantee optimality due to the uncertainty in $k_{lp}$. Search approaches, such as gradient-based  methods, can be used but may converge to a local optimum. In the case of a single element, we can use an exhaustive search to evaluate performance.

\end{Corollary}

\begin{figure}
	\centering 
	\includegraphics[width=3.7in]{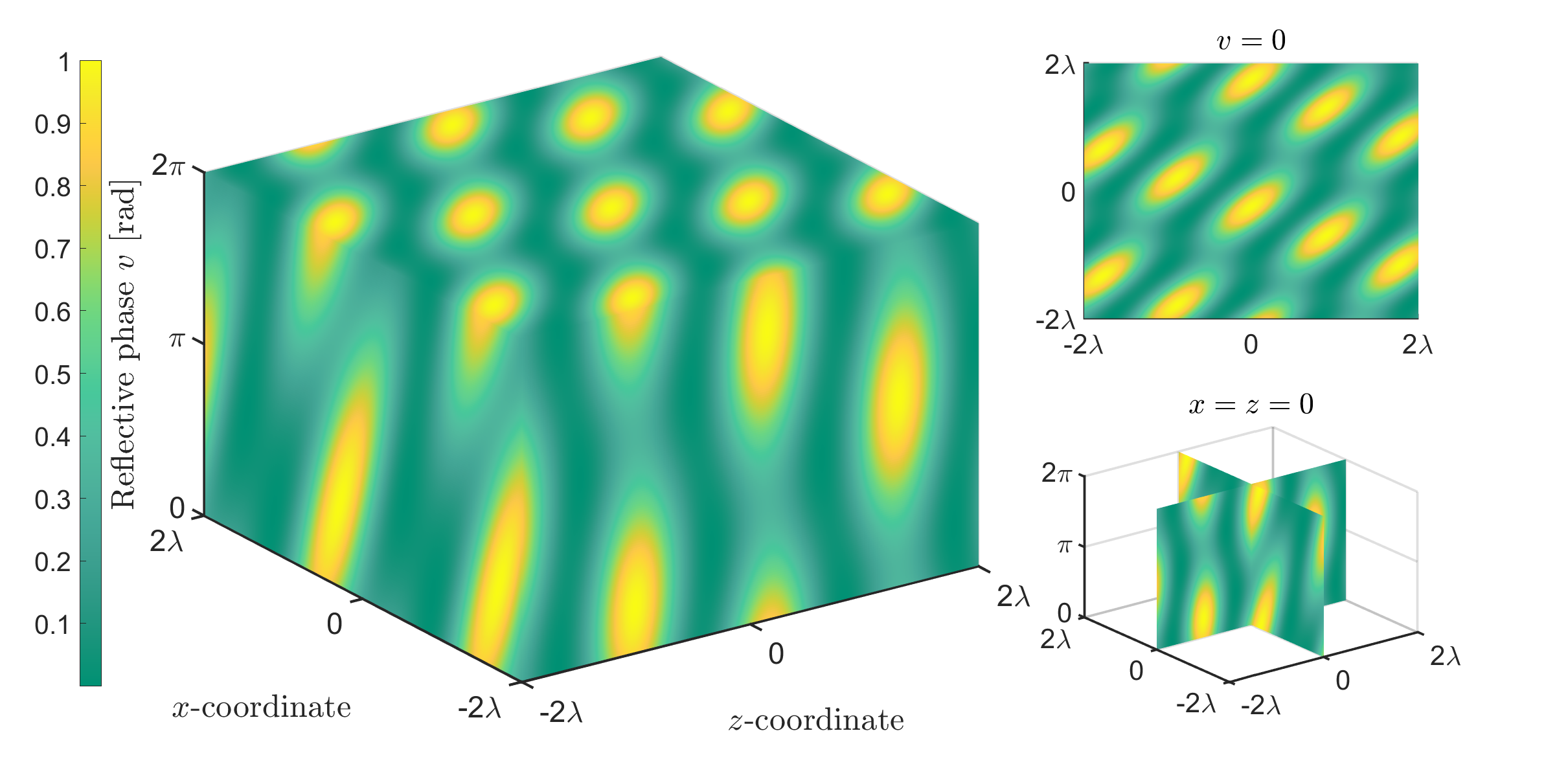}
	\caption{{Spatial} constructive/destructive interference with respect to $\{v,x,z\}$ in the single-element multi-path case.}\label{S-MP} 
\end{figure}

As shown in Fig. \ref{S-MP}, where $\bm{\theta}_{\rm B}=\left[\frac{\sqrt{2}}{2},\frac{1}{2}\right]$, $\bm{\theta}_{\rm U}=\left[0,-1\right]$,
$\bm{\phi}_{\rm B}=\left[\frac{\sqrt{2}}{4},\frac{\sqrt{6}}{4}\right]$, $\bm{\phi}_{\rm U}=\left[-\frac{\sqrt{3}}{2},0\right]$, $\bm{\alpha}=\left[e^{j\frac{\pi}{6}},e^{j\frac{\pi}{3}} \right]$, $\bm{\beta}=\left[e^{j\frac{\pi}{4}},e^{j\frac{\pi}{2}} \right]$, and $\gamma=e^{j\frac{\pi}{4}}$, we illustrate the constructive and destructive interference with respect to $\{v, x, z\}$. On the left side of Fig. \ref{S-MP}, the interference pattern exhibits a periodic property similar to that in Fig. \ref{S-SP}, but with a circular shape in the element position domain. Notably, not all modes reach an optimum. For instance, in the PBF-only mode with $x=z=0$, as shown on the bottom-right of Fig. \ref{S-MP}, the intersection line of two slices represents the objective function values as $v$ ranges from $0$ to $2\pi$. This line only encompasses sub-optimal values. This underscores the importance of the element position.

To gain further insights, the following proposition compares the EM-only and PBF-only modes for $LP=2$.
\begin{Proposition}\label{mi}
When the number of cascaded paths is set to $2$, i.e., $LP=2$, and $L=1$ and $P=2$ are used without loss of generality, a unique solution for $\{x,z\}$ can be found according to  (\ref{sqs}). In this case, this EM-only solution, obtained without optimizing $v$, results in a larger objective value compared to the PBF-only solution using  (\ref{vs}) while keeping $\{x,z\}$ fixed.
\end{Proposition}

\begin{Proof}
	For PBF-only case, substituting  (\ref{vs}) into (\ref{FBF}) yields the optimal PBF-only objective:
	\begin{equation}
		\begin{aligned}
		f_{\rm PBF}(v^\star) & =
		\left(  \vert h_d\vert+\vert h_{\rm cas}\vert \right)^2 \\  &=\left( \vert \gamma\vert +\frac{1}{\sqrt{2}}\left\vert   \alpha_1\beta^*_1+\alpha_1\beta^*_2    \right\vert\right)^2,
		\end{aligned}
	\end{equation} 
where $v^\star$ is obtained by  (\ref{vs}).

Solving the equation in  (\ref{sqs}) and combining with  (\ref{FEM}), we have
\begin{equation}\label{comp}
	\begin{aligned}
		f_{\rm EM}(x^\star,z^\star)= & \left\vert\gamma+     \frac{\alpha_1\beta^*_1}{\sqrt{2}}e^{j(\angle \gamma-\angle \alpha_1 \beta^*_1)}+\frac{\alpha_1\beta^*_2}{\sqrt{2}}e^{j(\angle \gamma-\angle \alpha_1 \beta^*_2)} \right\vert^2\\
		=& \left\vert  e^{j\angle\gamma} \left(\vert\gamma\vert+   \frac{1}{\sqrt{2}}\vert\alpha_1\beta^*_1\vert +\frac{1}{\sqrt{2}}\vert\alpha_1\beta^*_2 \vert \right)\right\vert^2\\
		=&   \left(\vert\gamma\vert + \frac{1}{\sqrt{2}}\vert\alpha_1\beta^*_1\vert +\frac{1}{\sqrt{2}}\vert\alpha_1\beta^*_2 \vert\right)^2 \\
		 \overset{(b)}{\geq}& f_{\rm PBF}(v^\star), 
	\end{aligned}
\end{equation}
where $\{x^\star,z^\star\}$ are obtained by solving  (\ref{sqs}), and $(b)$ holds {with} equality when the two cascaded path gains $\alpha_1\beta_1^*$ and $\alpha_1\beta_2^*$ are phase-aligned.

\end{Proof}

 (\ref{comp}) reveals that EM can induce constructive interference between the cascaded and direct channel paths, whereas PBF can only achieve constructive interference between the direct channel path and the entire cascaded channel. This is because PBF affects all cascaded paths uniformly.

Proposition \ref{mi} indicates that EM-only FIM can sometimes replace and even outperform PBF-only FIM. This finding is significant for practical implementation, suggesting that FIM, which only requires adjustments to the array structure like EM in this study, or other array shape adjustments, could substitute for PBF-based FIM. This approach could potentially reduce power consumption and hardware costs. However, in more complex scenarios, EM-only FIM may not deliver optimal performance. In such cases, combining EM with PBF for FIM may be necessary to achieve high performance. Importantly, the use of EM could allow for the adoption of discrete phases for PBF, offering a cost-effective and practical implementation option.

The following proposition provides a performance analysis for the single-element multi-path case with general values of $L$ and $P$.
\begin{Proposition}\label{ta1}
Assuming $\gamma \sim \mathcal{CN}(0, \sigma^2_\gamma)$, $\alpha_l \sim \mathcal{CN}(0, \sigma^2_\alpha)$, and $\beta_p \sim \mathcal{CN}(0, \sigma^2_\beta)$, the theoretical maximal received power for PBF-only FIM and the theoretical upper bound for FIM in the single-element multi-path case are, respectively,
\begin{equation}\label{t_bf}
\mathbb{E}\left\{ f_{\rm PBF}(v^\star)
\right\}=\sigma_\gamma^2+\frac{\pi\sqrt{\pi}}{4}\sigma_\gamma\sigma_\alpha\sigma_\beta
+	\sigma_\alpha^2\sigma_\beta^2,
\end{equation}
\begin{equation}\label{t_up}
	\begin{aligned}
		\mathbb{E}\left\{f_{\rm UB}\right\}=&\sigma_\gamma^2 + \frac{\pi\sqrt{\pi}\sqrt{LP}}{4} \sigma_\gamma\sigma_\alpha\sigma_\beta
		\\
		& +\left(1+\frac{ \pi (L-1) }{4}\right)\left(1+\frac{ \pi (P-1) }{4}\right)\sigma^2_\alpha\sigma^2_\beta,
	\end{aligned}
\end{equation}
where $	f_{\rm PBF}(v^\star)=\left(\vert\gamma\vert+\frac{1}{\sqrt{LP}} \left\vert  \sum_{l=1}^{L}\sum_{p=1}^{P} \alpha_l\beta_p^\star
\right\vert
\right)^2$ obtained by (\ref{FBF}) and (\ref{vs}) and $	f_{\rm UB} =\left(|\gamma|+ \frac{1}{\sqrt{LP}} \sum_{l=1}^{L}\sum_{p=1}^{P}|\alpha_l\beta^*_p| \right)^2$ obtained by  (\ref{hcd2}).
\end{Proposition}

\begin{Proof}
\textcolor{red}{See} the proof in Appendix \ref{A1}.
\end{Proof}

In the single-element multi-path case, as described in  (\ref{t_bf}), the maximal received power achieved by PBF-only FIM is independent of the number of cascaded paths, relying solely on the path gain. Conversely, the upper bound of FIM, given in  (\ref{t_up}), which \textcolor{red}{is} achieved by both EM-only and EM-PBF cases, indicates that PBF-only FIM can reach this upper bound only when $P=L=1$. In other cases, the PBF-only approach may perform significantly worse than the upper bound due to its neglect of multi-path interference.

\subsection{Multi-Element Multi-Path}

In the multi-element multi-path case, the cascaded channel is expressed as
$	h_{\rm cas}=  \sum_{l=1}^{L}\sum_{p=1}^{P} \alpha_l\beta^*_p \sum_{n=1}^N e^{jv_n} e^{j\frac{2\pi}{\lambda}\left(\widetilde{\theta}_{l,p}z_n+\widetilde{\phi}_{l,p}x_n\right)}$, yielding the objective:
\begin{equation}\label{hcd3}
	\begin{aligned}
&	f(\mathbf{v},\mathbf{x},\mathbf{z})\\
=& \left\vert\frac{1}{\sqrt{LP}}  \sum_{l=1}^{L}\sum_{p=1}^{P} \alpha_l\beta^*_p \sum_{n=1}^N e^{jv_n} e^{j\frac{2\pi}{\lambda}\left(\widetilde{\theta}_{l,p}z_n+\widetilde{\phi}_{l,p}x_n\right)}+\gamma \right\vert^2.
	\end{aligned}
\end{equation}  

\begin{Proposition}\label{MMF}
Considering the allowed inter-element spacing and feasible position in FIM, we can formulate the optimization problem to maximize the objective in (\ref{hcd3}):
\begin{equation}
	\begin{aligned}
		&\underset{\mathbf{v},\mathbf{x},\mathbf{z}}{\rm arg \ max} \ f (\mathbf{v},\mathbf{x},\mathbf{z})\\
		&{\rm s.t.} \ \Vert\mathbf{p}_{n_1}-\mathbf{p}_{n_2}\Vert_2 \geq d_{\rm min} ,\\
	& \ \ \ \ \ \	\mathbf{p}_{n_1},\mathbf{p}_{n_2} \in \mathcal{R} ,
		\\
		& \ \ \ \ \ \forall n_1,n_2 \in\{1,\cdots,N\}, n_1\neq n_2,
	\end{aligned}
\end{equation}
where $\mathbf{p}_n\triangleq[x_n,z_n]^T$ denotes the coordinate of the $n$-th FIM element
, $d_{\rm min}$ is the allowed minimal inter-element spacing that avoids severe mutual coupling effects.

 This problem can be addressed using a constraint gradient-based search method; however, due to its local optimization nature, it may not achieve optimal performance. In this study, global optimization is employed to fully exploit the potential of FIM, as detailed in Section \ref{BO}
\end{Proposition}

\begin{Corollary}
	Considering the PBF-only mode for FIM, the optimization problem is given by
	\begin{equation}
		\begin{aligned}
			\underset{\mathbf{v}}{\rm arg \ max} \ f_{\rm PBF}(\mathbf{v}) ,
		\end{aligned}
	\end{equation}
	where
	\begin{equation}\label{fBFMM}
		\begin{aligned}
			f_{\rm PBF}(\mathbf{v})\triangleq \left\vert \frac{1}{\sqrt{LP}}  \sum_{l=1}^{L}\sum_{p=1}^{P} \alpha_l\beta^*_p \sum_{n=1}^N e^{jv_n}+\gamma \right\vert^2.
		\end{aligned}
	\end{equation} 
	Consider the phase alignment strategy for each $v_n$, this translates into addressing
	\begin{equation}
		\angle\left(\sum_{n=1}^N e^{jv_n}\right)=	\angle \gamma -\angle \left(\sum_{l=1}^{L}\sum_{p=1}^{P} \alpha_l\beta^*_p\right).
	\end{equation}
	By setting $v_1=\cdots=v_N$, a solution can be derived by, $\forall n$:
	\begin{equation}\label{vMM}
		v_n=\angle \gamma -\angle \left(\sum_{l=1}^{L}\sum_{p=1}^{P} \alpha_l\beta^*_p\right) .
	\end{equation}
\end{Corollary}
\begin{Corollary}
	Considering the EM-only mode for FIM, the optimization problem is given by 
\begin{equation}
	\begin{aligned}
		&\underset{\mathbf{x},\mathbf{z}}{\rm arg \ max} \ f_{\rm EM}(\mathbf{x},\mathbf{z})\\
		&{\rm s.t.} \ \Vert\mathbf{p}_{n_1}-\mathbf{p}_{n_2}\Vert_2 \geq d_{\rm min} ,\\ 
			& \ \ \ \ \ \	\mathbf{p}_{n1},\mathbf{p}_{n2} \in \mathcal{R} ,
		\\
		& \ \ \ \ \ \forall n_1,n_2 \in\{1,\cdots,N\}, n_1\neq n_2,
	\end{aligned}
\end{equation}
where 
\begin{equation}
f_{\rm EM}(\mathbf{x},\mathbf{z})\triangleq \frac{1}{\sqrt{LP}}  \sum_{l=1}^{L}\sum_{p=1}^{P} \alpha_l\beta^*_p \sum_{n=1}^N  e^{j\frac{2\pi}{\lambda}\left(\widetilde{\theta}_{l,p}z_n+\widetilde{\phi}_{l,p}x_n\right)}+\gamma.
\end{equation} 
 
Similar to Proposition \ref{MMF}, the problem of element position optimization is difficult to solve with closed-form solutions in the multi-element multi-path case. Therefore, Bayesian optimization in Section \ref{BO} is used to evaluate the EM-only case.
\end{Corollary}

The following proposition provides a performance analysis for the multi-element multi-path case with general values of $L$ and $P$.
\begin{Proposition}\label{ta2}
Assuming $\gamma \sim \mathcal{CN}(0, \sigma^2_\gamma)$, $\alpha_l \sim \mathcal{CN}(0, \sigma^2_\alpha)$, and $\beta_p \sim \mathcal{CN}(0, \sigma^2_\beta)$, the theoretical maximal received power for PBF-only FIM and the theoretical upper bound for FIM in the multi-element multi-path case are, respectively,
\begin{equation} 
	\mathbb{E}\left\{ f_{\rm PBF}(v^\star)
	\right\}=\sigma_\gamma^2+\frac{\pi\sqrt{\pi}N}{4}\sigma_\gamma\sigma_\alpha\sigma_\beta
	+N^2	\sigma_\alpha^2\sigma_\beta^2,
\end{equation}
\begin{equation} 
	\begin{aligned}
		\mathbb{E}\left\{f_{\rm UB}\right\}=&\sigma_\gamma^2 + \frac{\pi\sqrt{\pi}\sqrt{LP} N}{4} \sigma_\gamma\sigma_\alpha\sigma_\beta
		\\
		& +N^2\left(1+\frac{ \pi (L-1) }{4}\right)\left(1+\frac{ \pi (P-1) }{4}\right)\sigma^2_\alpha\sigma^2_\beta,
	\end{aligned}
\end{equation}
where $	f_{\rm PBF}(v^\star)$ is obtained by  (\ref{fBFMM}) and (\ref{vMM}) and $	f_{\rm UB}$ is obtained by  (\ref{hcd3}).
\end{Proposition}
\begin{Proof}
This proof can be derived similar to  Appendix \ref{A1}.
\end{Proof}

Compared to Proposition \ref{ta1}, we can see that the theoretical bound has a squared scaling law with the number of FIM elements.

\subsection{Bayesian Optimization for EM-PBF}\label{BO}

{Due to the difficulty in obtaining closed-form expressions for EM and PBF parameters in the multi-path scenario, as described in Sections III-C and III-D, we utilize Bayesian optimization, a global optimization method, to maximize the received power.} 
Bayesian Optimization consists of two components: the surrogate model and the acquisition function. In this study, the Gaussian process is used as the surrogate model, and the expected improvement (EI) acquisition function is used. The Gaussian process predicts the posterior probability based on the measured function values, while the EI acquisition function provides suggestions on which variable values to measure in the next iteration, considering variable constraints. When the most promising variable value is obtained by the EI strategy, it is used to calculate the objective function, yielding a measured function value that updates the posterior probability. This iterative process continues to optimize the objective function progressively.
Several studies have utilized Bayesian optimization for communication areas, including beam alignment/training \cite{BO1} and edge computing \cite{BO2}. However, in our case, the inter-element spacing constraint should be incorporated into the Bayesian optimization process. This is feasible thanks to the advancements in constrained Bayesian optimization research \cite{BO3}. 

\begin{figure}
	\centering
	\subfigure[$N=1$.]{
		\includegraphics[width=2.8in]{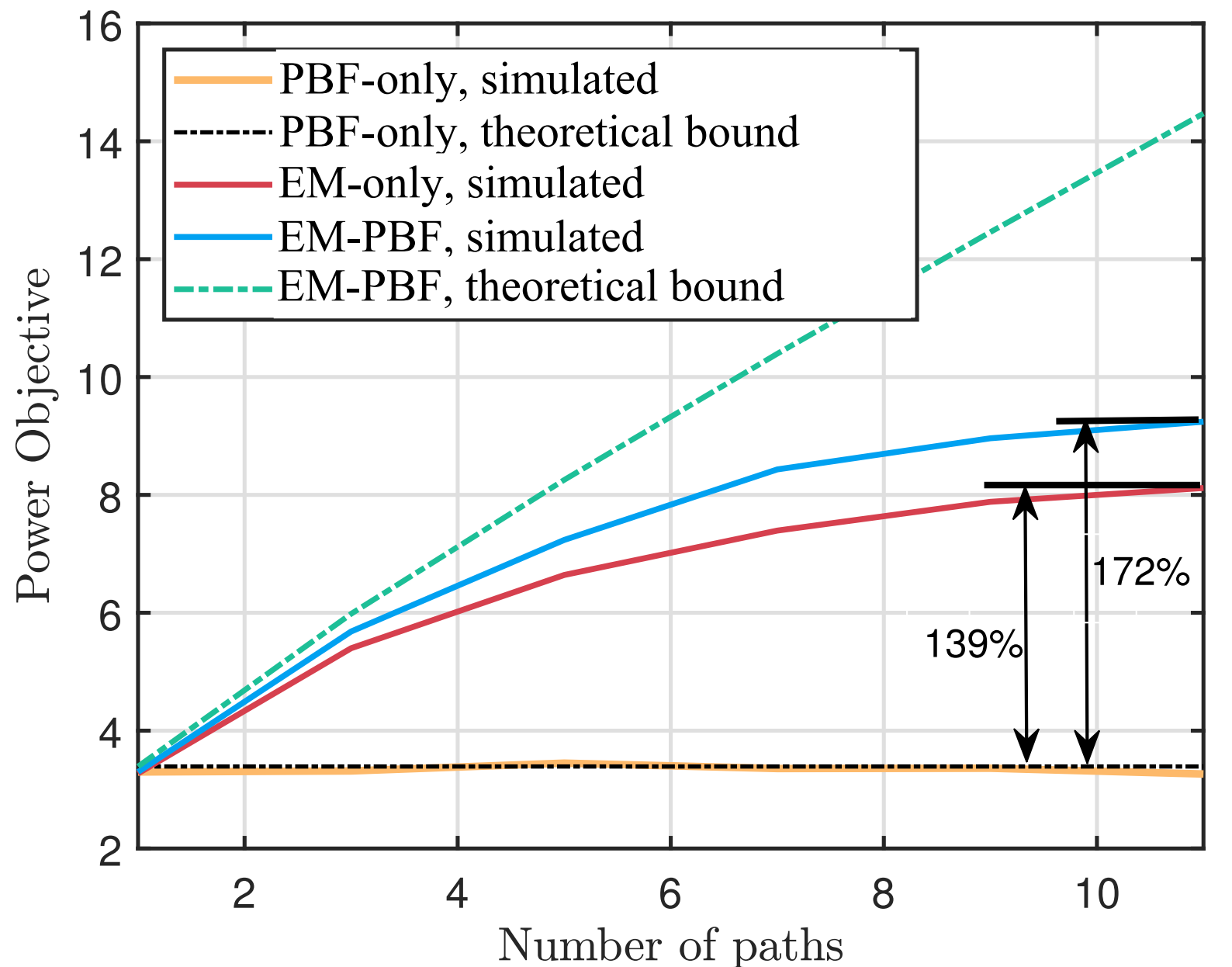}
	}
	\\    
	\subfigure[$N=2$.]{
		\includegraphics[width=2.7in]{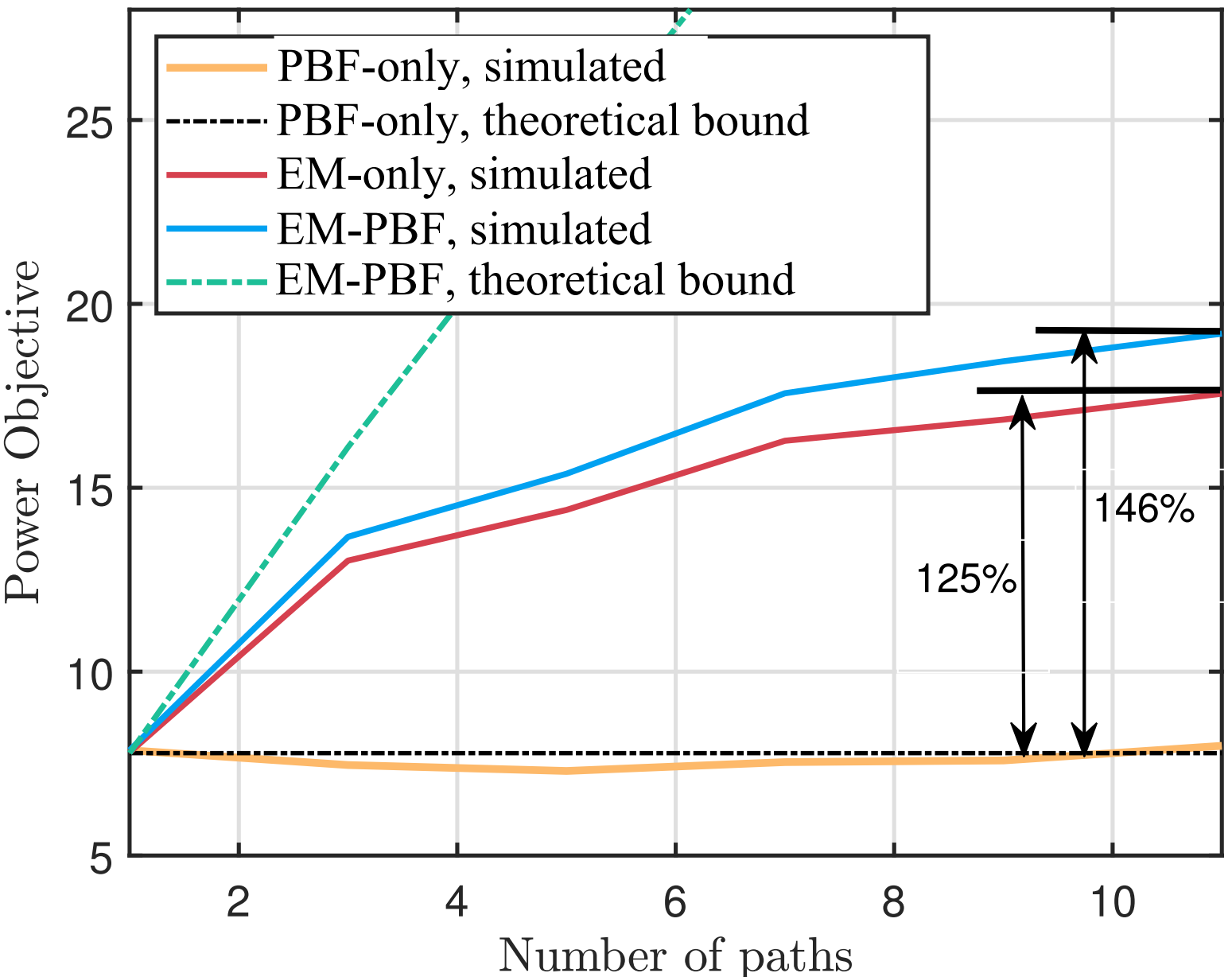}
	} 
	\caption{The power objective value of different modes versus the number of paths in the multi-path case.}
	\label{M-MP}
\end{figure}
As shown in Fig. \ref{M-MP}, we evaluate the PBF-only, EM-only, and EM-PBF modes, with $\sigma_\alpha=\sigma_\beta=\sigma_\gamma=1$, $L=1$, $P$ varying from 1 to 12, and the movable region $R$ set to $\lambda$, constraining the elements within a $\lambda\times \lambda$ square. Additionally, `simulated' refers to the solutions for objective maximization, such as the closed-form expression for PBF-only and Bayesian optimization for the EM-only and EM-PBF modes, while `theoretical' indicates the derived bounds in Propositions \ref{ta1} and \ref{ta2}. Figs. \ref{M-MP}(a) and \ref{M-MP}(b) depict the power maximization for different modes under single- and multi-element FIM configurations. It is evident that EM-only and EM-PBF outperform PBF-only, as the latter shows no performance gain with an increasing number of paths. The figures also reveal a gap between the derived upper bound and the simulated mode, suggesting room for algorithmic improvement.
 
{In this section, we assume that the channel parameters used for optimizing EM and PBF are perfect. The method for estimating the channel in the FIM system is detailed in Section IV.}
\section{Channel Estimation for FIM}
The previous received power maximization problem relies on accurate channel information, including path angle and path gain information. Compared to spatial channel estimation for traditional RIS, FIM considers both PBF and EM.

\subsection{Estimation Framework}
\begin{figure}
	\centering 
	\includegraphics[width=3.4in]{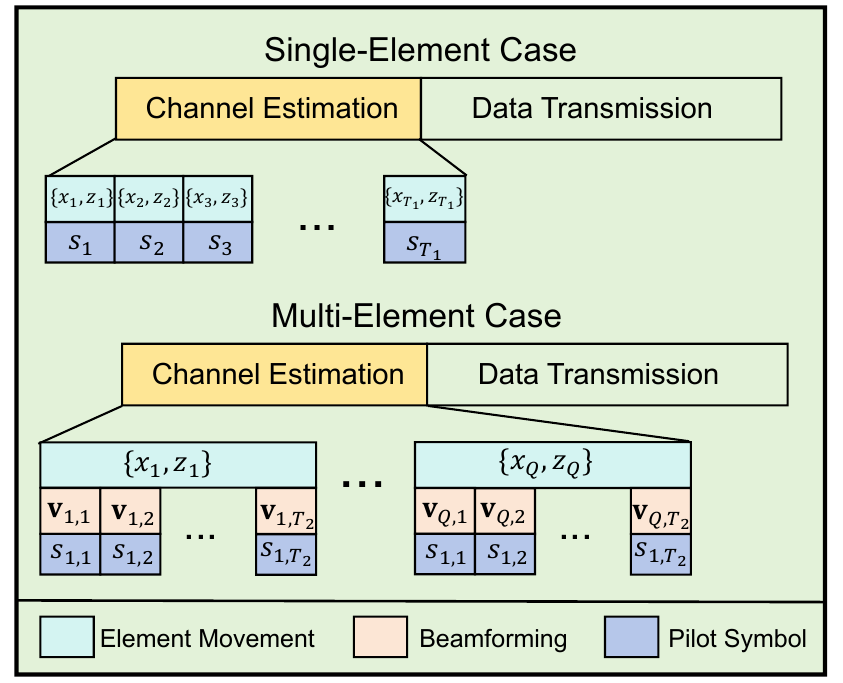}
	\caption{Channel estimation protocol for FIM.}\label{protocol} 
\end{figure}

We propose two estimation protocols for single-element and multi-element FIM, respectively:
 
 \begin{itemize}
 	\item	Single-Element FIM: In each time slot, the element moves once, and the receiver collects the transmitted signal. With $T_1$ time slots, these $T_1$ movements form a virtual FIM for spatial prameter estimation. This process is shown in Fig. \ref{protocol}.
 	
 	\item	Multi-Element FIM: Assuming $Q$ subframes with $T_2$ time slots in each subframe,  the element moves once per subframe, and the FIM adjusts the phase once per time slot. The total number of time slots is $QT_2$. This process is shown in Fig. \ref{protocol}.
 \end{itemize}

\subsubsection{Single-Element FIM} 
The received pilot signal in the $t$-th time slot, $t\in\{1,\cdots,T_1\}$, is given by
\begin{equation}
	\begin{aligned}
		y_t=&\left(e^{j{v}_t}{\rm diag}({h}_{\rm FU}^H) {h}_{\rm BF} +h_d\right)s_t+n_t  \\
		=&\frac{1}{\sqrt{LP}}e^{j{v}_t}s_t\sum_{l=1}^{L}\sum_{p=1}^P\alpha_l \beta^*_p e^{j\frac{2\pi}{\lambda}(\widetilde{\theta}_{l,p}z_t +
			\widetilde{\phi}_{l,p}x_t)}+ h_ds_t+n_t.
	\end{aligned}
\end{equation}

By setting $s_t=1$ and $e^{jv_t}=1$ for all $t$, we stack $\{y_t\}_{t=1}^{T_1}$ into $\mathbf{y}\in\mathbb{C}^{T_1\times 1}$, given by
\begin{equation}\label{yb}
\mathbf{y}= \mathbf{A}\bm{\eta}+\bm{\gamma}+\mathbf{n},
\end{equation}
where $\mathbf{A}\triangleq \left[\mathbf{a}_{T_1}\left(\widetilde{\theta}_1,\widetilde{\phi}_1\right),\cdots,\mathbf{a}_{T_1}\left(\widetilde{\theta}_{LP},\widetilde{\phi}_{LP}\right)\right]\in\mathbb{C}^{T_1\times LP}$,
\begin{equation}\label{a}
\mathbf{a}_{T_1}\left(\widetilde{\theta},\widetilde{\phi}\right)\triangleq \left[e^{j\frac{2\pi}{\lambda}(\widetilde{\theta}z_1+\widetilde{\phi}x_1)}, \cdots, e^{j\frac{2\pi}{\lambda}(\widetilde{\theta}z_{T_1}+\widetilde{\phi}x_{T_1})}\right]^T,
\end{equation} 
$\bm{\eta}\triangleq \frac{1}{\sqrt{LP}}[\alpha_1\beta^*_1,\cdots, \alpha_L\beta^*_P]^T\in\mathbb{C}^{LP\times 1}$, $\bm{\gamma}\triangleq [\gamma,\cdots,\gamma]^T\in\mathbb{C}^{T_1\times 1}$ is the direct channel coefficient, and $\mathbf{n}\triangleq[n_1,\cdots,n_Q]^T\in\mathbb{C}^{T_1\times 1}$. 

Now, we can formulate a parameter estimation problem as
\begin{equation}\label{es1}
	\underset{\widetilde{\bm{\theta}},\widetilde{\bm{\phi}},\bm{\eta},\gamma}{\rm arg \min} \left\Vert \mathbf{y}-\mathbf{A}\bm{\eta}-\bm{\gamma} \right\Vert_2^2.
\end{equation}

{Note that} $\bm{\gamma}$ can be written as
\begin{equation}\label{ghj}
	\begin{aligned}
		\bm{\gamma}=&\gamma\mathbf{1} \\
		=& \gamma \mathbf{a}_{T_1}(0,0),
	\end{aligned}
\end{equation} which means that the direct channel serves a special path of the cascaded channel with $\widetilde{\theta}=\widetilde{\phi}=0$. 
Then, we can write  (\ref{yb}) as
\begin{equation}
	\mathbf{y}=\widetilde{\mathbf{A}} \widetilde{\bm{\eta}} +\mathbf{n},
\end{equation}
where $\widetilde{\mathbf{A}}\triangleq [\mathbf{a}_{T_1}(0,0), \mathbf{A}]\in\mathbb{C}^{T_1\times(LP+1)}$ and $\widetilde{\bm{\eta}}\triangleq \begin{bmatrix}
\gamma\\ \bm{\eta}
\end{bmatrix} \in\mathbb{C}^{LP+1}$.

Then, the problem in  (\ref{es1}) is re-written by
\begin{equation}\label{CS1}
	\underset{\widetilde{\bm{\theta}},\widetilde{\bm{\phi}},\widetilde{\bm{\eta}}}{\rm arg \min} \left\Vert \mathbf{y}-\widetilde{\mathbf{A}}\widetilde{\bm{\eta}} \right\Vert_2^2.
\end{equation}

{To solve the problem,}
we can establish a dictionary $\bm{\Phi}\in\mathbb{C}^{T_1\times G}$ containing atoms sampling $G$ different $\left\{\widetilde{\theta},\widetilde{\phi}\right\}$, where $\left\{\widetilde{\theta}=0,\widetilde{\phi}=0\right\}$ must be included for the direct channel. {This leads to a sparse recovery problem:}
\begin{equation}\label{sr1}
{\rm min}\ \left\Vert\bm{\xi}\right\Vert_0, \ {\rm s.t.} \ \left\Vert\mathbf{y}-\bm{\Phi}\bm{\xi}\right\Vert_2^2\leq \epsilon,
\end{equation}
where $\bm{\xi}\in\mathbb{C}^{G\times 1}$ is the $(LP+1)$-sparse signal in which each nonzero element denotes the path gain, and $\epsilon$ denotes the precise factor. 



\subsubsection{Multi-Element FIM}
In the $t$-the time slot of the $q$-th subframe, $t\in\{1,\cdots,T_2\}$, $q\in\{1,\cdots,Q\}$, the received signal is expressed by
\begin{equation}
	\begin{aligned}
		y_{q,t}=&\left(\mathbf{v}^T_{q,t}{\rm diag}(\mathbf{h}_{\rm FU}^H) \mathbf{h}_{\rm BF}+ h_d \right)s_{q,t}+n_{q,t} \\
		=& \frac{s_{q,t}}{\sqrt{LP}} \mathbf{v}^T_{q,t}\sum_{l=1}^{L}\sum_{p=1}^P\alpha_l \beta^*_p e^{j\frac{2\pi}{\lambda}(\widetilde{\theta}_{l,p}\mathbf{z}_q+\widetilde{\phi}_{l,p}\mathbf{x}_q)}+\gamma s_{q,t}+n_{q,t},
	\end{aligned}
\end{equation}
where $\{\mathbf{x}_q,\mathbf{z}_q\}$ dentoes the element positions of the $q$-th subframe, $\mathbf{v}_{q,t}\in\mathbb{C}^{N\times 1}$ denotes the PBF vector in the $t$-th time slot, and $n_{q,t}$ is the noise.

With signals collected in $T_2$ time slots, the column-stacked signal $\mathbf{y}_q\in\mathbb{C}^{T_2\times 1}$ in the $q$-th subframe is given by 
\begin{equation}
	\begin{aligned} 
	\mathbf{y}_q=&\frac{1}{\sqrt{LP}}\mathbf{W}_q^T \sum_{l=1}^{L}\sum_{p=1}^P\alpha_l \beta^*_p e^{j\frac{2\pi}{\lambda}(\widetilde{\theta}_{l,p}\mathbf{z}_q+\widetilde{\phi}_{l,p}\mathbf{x}_q)}+
	\bm{\gamma}+\mathbf{n}_q,
\end{aligned}
\end{equation}
where $\mathbf{W}_q\triangleq[\mathbf{v}_{q,1},\cdots,\mathbf{v}_{q,T_2}]\in\mathbb{C}^{N\times T_2}$, $\bm{\gamma}\triangleq [\gamma,\cdots,\gamma]^T \in\mathbb{C}^{T_2\times 1}$,
 and $\mathbf{n}_q\triangleq [n_{q,1},\cdots,n_{q,T_2}]\in\mathbb{C}^{T_2\times 1}$ is the stacked noise vector.

Finally, stacking all signals in $Q$ subframes by columns, we obtain
\begin{equation}
	\begin{aligned}
	&	\widetilde{\mathbf{y}}\\ \triangleq &\frac{1}{\sqrt{LP}} \begin{bmatrix}
	\mathbf{W}^T_1 \sum_{l}^{L}\sum_{p}^P\alpha_l \beta^*_p e^{j\frac{2\pi}{\lambda}(\widetilde{\theta}_{l,p}\mathbf{z}_1+\widetilde{\phi}_{l,p}\mathbf{x}_1)} \\ \vdots \\	\mathbf{W}^T_Q \sum_{l}^{L}\sum_{p}^P\alpha_l \beta^*_p e^{j\frac{2\pi}{\lambda}(\widetilde{\theta}_{l,p}\mathbf{z}_Q+\widetilde{\phi}_{l,p}\mathbf{x}_Q)}
		\end{bmatrix}+\widetilde{\bm{\gamma}} +\widetilde{\mathbf{n}}
	 \\
	=& \frac{1}{\sqrt{LP}}\begin{bmatrix}
	\mathbf{W}^T_1& & 
\\
&	\ddots&\\
& &\mathbf{W}^T_Q
	\end{bmatrix} \begin{bmatrix}
\sum_{l}^{L}\sum_{p}^P\alpha_l \beta^*_p e^{j\frac{2\pi}{\lambda}(\widetilde{\theta}_{l,p}\mathbf{z}_1+\widetilde{\phi}_{l,p}\mathbf{x}_1)} \\ \vdots \\ \sum_{l}^{L}\sum_{p}^P\alpha_l \beta^*_p e^{j\frac{2\pi}{\lambda}(\widetilde{\theta}_{l,p}\mathbf{z}_Q+\widetilde{\phi}_{l,p}\mathbf{x}_Q)}
\end{bmatrix}  \\
&+\widetilde{\bm{\gamma}} +\widetilde{\mathbf{n}}\\
=& \frac{1}{\sqrt{LP}}\widetilde{\mathbf{W}}^T \sum_{l=1}^{L}\sum_{p=1}^P\alpha_l \beta^*_p e^{j\frac{2\pi}{\lambda}(\widetilde{\theta}_{l,p}\widetilde{\mathbf{z}}+\widetilde{\phi}_{l,p}\widetilde{\mathbf{x}})}+\widetilde{\bm{\gamma}} +\widetilde{\mathbf{n}},
	\end{aligned}
\end{equation}
where $\widetilde{\mathbf{W}}\triangleq\begin{bmatrix}
	\mathbf{W}_1& & 
	\\
	&	\ddots&\\
	& &\mathbf{W}_Q
\end{bmatrix}\in\mathbb{C}^{NT_2\times QT_2}$, $\widetilde{\mathbf{x}}\triangleq [\mathbf{x}_1^T,\cdots,\mathbf{x}_Q^T]^T\in\mathbb{C}^{NT_2\times 1}$, $\widetilde{\mathbf{z}}\triangleq [\mathbf{z}_1^T,\cdots,\mathbf{z}_Q^T]^T\in\mathbb{C}^{NT_2\times 1}$, $\widetilde{\gamma}\triangleq [\bm{\gamma}^T,\cdots,\bm{\gamma}^T]^T\in\mathbb{C}^{QT_2\times 1}$, and $\widetilde{\mathbf{n}}\triangleq[\mathbf{n}_1^T,\cdots,\mathbf{n}_Q^T]^T\in\mathbb{C}^{QT_2\times 1}$.

Similar to  (\ref{yb}), the above equation can be more compact:
\begin{equation}\label{yb2}
	\widetilde{\mathbf{y}} =\widetilde{\mathbf{W}}^T \mathbf{A}\bm{\eta}+\widetilde{\bm{\gamma}} +\widetilde{\mathbf{n}},
\end{equation}
where $\mathbf{A}\triangleq \left[\mathbf{a}_{NT_2}\left(\widetilde{\theta}_1,\widetilde{\phi}_1\right),\cdots,\mathbf{a}_{NT_2}\left(\widetilde{\theta}_{LP},\widetilde{\phi}_{LP}\right)\right]\in\mathbb{C}^{NT_2\times LP}$,
\begin{equation}\label{a4}
	\mathbf{a}_{NT_2}\left(\widetilde{\theta},\widetilde{\phi}\right) \triangleq \left[e^{j\frac{2\pi}{\lambda}(\widetilde{\theta}z_1+\widetilde{\phi}x_1)}, \cdots, e^{j\frac{2\pi}{\lambda}(\widetilde{\theta}z_{NT_2}+\widetilde{\phi}x_{NT_2})}\right]^T,
\end{equation} 
and $\bm{\eta}\triangleq\sqrt{\frac{1}{LP}} [\alpha_1\beta^*_1,\cdots, \alpha_L\beta^*_P]^T\in\mathbb{C}^{LP\times 1}$. 

Similar to  (\ref{ghj}), we regard the direct channel component $\widetilde{\bm{\gamma}}$ as $\gamma	\mathbf{a}_{QT_2}\left(0,0\right)$, and re-write  (\ref{yb2}) by
\begin{equation}
	\begin{aligned}
		\widetilde{\mathbf{y}}&=\widetilde{\mathbf{W}}^T {\mathbf{A}}\bm{\eta}+\mathbf{a}_{QT_2}(0,0)\gamma+\widetilde{\mathbf{n}} \\
		&= \widetilde{\mathbf{A}}\widetilde{\bm{\eta}}+\widetilde{\mathbf{n}}
	\end{aligned}
\end{equation}
  where $\widetilde{\mathbf{A}}\triangleq \left[\mathbf{a}_{QT_2}(0,0), \widetilde{\mathbf{W}}^T\mathbf{A}\right]\in\mathbb{C}^{QT_2\times(LP+1)}$ and $\widetilde{\bm{\eta}}\triangleq \begin{bmatrix}
  	\gamma\\ \bm{\eta}
  \end{bmatrix} \in\mathbb{C}^{LP+1}$.
  Then, yielding the same form in  (\ref{CS1}):
  \begin{equation}\label{CS2}
  	\underset{\widetilde{\bm{\theta}},\widetilde{\bm{\phi}},\widetilde{\bm{\eta}}}{\rm arg \min} \left\Vert \widetilde{\mathbf{y}}- \widetilde{\mathbf{A}}\widetilde{\bm{\eta}}\right\Vert_2^2.
  \end{equation}
 
 Similar to  (\ref{sr1}), we establish a dictionary $\bm{\Phi}\in\mathbb{C}^{QT_2\times G}$ according to $\widetilde{\mathbf{A}}$.  Generating $\bm{\Phi}$ here is slightly different from below  (\ref{CS2}) due to the measurement matrix.  Given the dictionary $\bm{\Phi}$, we attain
 \begin{equation}\label{sr2}
 	{\rm min}\ \left\Vert\bm{\xi}\right\Vert_0, \ {\rm s.t.} \ \left\Vert\mathbf{y}-\bm{\Phi}\bm{\xi}\right\Vert_2^2\leq \epsilon,
 \end{equation}
 where $\bm{\xi}\in\mathbb{C}^{G\times 1}$ and $\epsilon$ share same definitions below  (\ref{sr1}).  
 
 Therefore, we find that channel estimation for both single-element and multi-element FIM can be formulated as a standard CS problem {with different sensing matrices.}
   
\subsection{Recovery Algorithm}\label{RA}

The proposed channel estimation frameworks in  (\ref{sr1}) and (\ref{sr2}) {can be addressed using standard sparse recovery algorithms \cite{eb}}, which have been extensively studied across various methodologies, including greedy iteration, convex optimization, message passing, Bayesian learning, and deep learning.

Before presenting the proposed CMFB-SBL algorithm, we first introduce the two-layer hierarchical prior-based SBL model for the linear form
$
	\mathbf{y}=\bm{\Phi}\bm{\xi}+\mathbf{n}$.
The two-layer hierarchical prior that promotes sparsity is commonly used for $\bm{\xi}\triangleq [\xi_1,\cdots,\xi_{G}]^T$. Specifically, $\bm{\xi}$ is assumed to follow a complex Gaussian distribution parameterized by $\bm{\rho}\triangleq [\rho_1,\cdots,\rho_G]^T$, with $\rho_i$ being the inverse variance of $\xi_i$, $\forall i\in\{1,\cdots,G\}$:
\begin{equation}
	p(\bm{\xi} | \bm{\rho})=\prod_{i=1}^G p\left(\xi_i | \rho_i\right)=\prod_{i=1}^G \mathcal{CN}\left(0, \rho_i^{-1}\right) .
\end{equation}

Further, a Gamma prior is considered over $\bm{\rho}$:
\begin{equation}
	p(\boldsymbol{\rho}|a,b)=\prod_{i=1}^G p\left(\rho_i | a, b\right)=\prod_{i=1}^G \Gamma^{-1}(a) b^a \rho_i^{a-1} e^{-b \rho_i} .
\end{equation}

By marginalizing over the hyperparameters $\bm{\rho}$, the overall prior on $\bm{\xi}$ is then evaluated as
\begin{equation}
	p(\bm{\xi}|a,b)=\prod_{i=1}^N \int_0^{\infty} \mathcal{CN}\left(\xi_i | 0, \rho_i^{-1}\right) \Gamma\left(\rho_i | a, b\right) {\rm d} \rho_i .
\end{equation}

The inverse of noise variance $\tau$ is also assumed to follow a Gamma prior, $p(\tau)=\Gamma^{-1}(c) d^c \tau^{c-1} e^{-d \tau}$. Now the likelihood distribution can be written as
\begin{equation}
	p(\mathbf{y} |\bm{\xi}, \sigma^2)=(2 \pi\sigma^2)^{-G / 2}   e^{\frac{- \|\mathbf{y}-\bm{\Phi} \bm{\xi}\|^2}{2 \sigma^2}} . 
\end{equation}

\subsubsection{V-SBL}
The variational framework typically tackles inference models with analytically intractable evidence, similar to the goal of sampling methods. Variational methods solve this problem by introducing a distribution $q$, which splits the log evidence into two terms:
\begin{equation}
	\log p(\mathbf{y})=\int q(\bm{\Theta})\log \frac{p(\mathbf{y},\bm{\Theta})}{q(\bm{\Theta})}-\int q(\bm{\Theta})\log \frac{p(\bm{\Theta}|\mathbf{y})}{q(\bm{\Theta})},
\end{equation}
 where $\bm{\Theta}\triangleq [\bm{\xi},\bm{\rho},\sigma^2]$ denotes the parameter of interest, and
 the former term is evidence lower bound (ELBO) and the latter is Kullback-Leibler (KL) divergency. As $\log p(\mathbf{y})$ is irrelated to $q(\bm{\Theta})$, maximizing ELBO is equivalent to minimizing KL divergency. This promotes $q(\bm{\Theta})$ approaching $p(\bm{\Theta}|\mathbf{y})$ when the KL divergency gets minimum.
 
Adopting a a factorized form over $\bm{\Theta}$ \cite{vsbl}, such that
\begin{equation}\label{qt}
	q(\bm{\Theta})=q(\bm{\xi})q(\bm{\rho})q(\sigma^2),
\end{equation}
the ELBO can then be maximized over all possible factorial distributions by performing a free-form
maximization over $\{\bm{\xi},\bm{\rho},\sigma^2\}$ alternatively, yielding the optimal $q(\bm{\xi})$, $q(\bm{\rho})$, $q(\sigma^2)$:
\begin{equation}\label{qx}
q(\bm{\xi})=\frac{e^{\mathbb{E}_{\bm{\rho},\sigma^2} \{\log p(\mathbf{y},\bm{\Theta})\}}}{\int e^{\mathbb{E}_{\bm{\rho},\sigma^2} \{\log p(\mathbf{y},\bm{\Theta})\}} {\rm d} \bm{\xi}},
\end{equation}
\begin{equation}\label{qr}
	q(\bm{\rho})=\frac{e^{\mathbb{E}_{\bm{\xi},\sigma^2} \{\log p(\mathbf{y},\bm{\Theta})\}}}{\int e^{\mathbb{E}_{\bm{\xi},\sigma^2} \{\log p(\mathbf{y},\bm{\Theta})\}} {\rm d} \bm{\rho}},
\end{equation}
\begin{equation}\label{qs}
	q(\sigma^2)=\frac{e^{\mathbb{E}_{\bm{\xi},\bm{\rho}} \{\log p(\mathbf{y},\bm{\Theta})\}}}{\int e^{\mathbb{E}_{\bm{\xi},\bm{\rho}} \{\log p(\mathbf{y},\bm{\Theta})\}} {\rm d} \sigma^2},
\end{equation}
where \begin{equation}\label{pyt}
	\begin{aligned}
		&	\log p(\mathbf{y},\bm{\Theta})=p(\mathbf{y}|\bm{\xi},\bm{\rho},\sigma^2)p(\bm{\xi}|\bm{\rho})p(\bm{\rho})p(\sigma^2)\\
	  \  &= -\frac{G}{2}\log \sigma^2 -\frac{1}{2\sigma^2} \Vert \mathbf{y}-\bm{\Phi}\bm{\xi}\Vert^2_2 \\
		& \ +\sum_{i=1}^{G} \left( \frac{1}{2}\log \rho_i-\frac{\rho_i}{2}\vert\xi_i\vert^2 +(a-1)\log\rho_i+a\log b-b\rho_i
		\right)\\
		& \ -(c-1)\log\sigma^2+c\log d-\frac{d}{\sigma^2}+Const.
	\end{aligned}
\end{equation}

Combining  (\ref{qx}) and (\ref{pyt}), we can obtain mean $\bm{\mu}$ and covariance $\bm{\Sigma}$ of $q(\bm{\xi})$:
\begin{equation}
	\bm{\mu}=\mathbb{E}\{\sigma^{-2}\}\bm{\Sigma}\bm{\Phi}^H\mathbf{y},
\end{equation}
\begin{equation}\label{var2}
	\bm{\Sigma}=\left( \mathbb{E}\{\sigma^{-2}\}\bm{\Phi}^H\bm{\Phi}+ {\rm diag}(\bm{\rho})
	\right)^{-1}.
\end{equation} 

Meanwhile, according to (\ref{qr}), (\ref{qs}), and (\ref{pyt}), we can get the parameter updating rule:
\begin{equation}\label{rho2}
	\mathbb{E}\{\rho_i\}=\frac{a+\frac{1}{2}}{\frac{\vert\mu_i\vert^2+[\bm{\Sigma}]_{i,i}}{2}+b},
\end{equation}
\begin{equation}\label{s2}
\mathbb{E}\{\sigma^{-2}\}=\frac{c+\frac{G}{2}}{\frac{\Vert\mathbf{y}\Vert^2_2-2\Re\{\mathbf{y}^H\bm{\Phi}\bm{\mu}\}+{\rm Tr}\left(\bm{\Phi}^H\bm{\Phi}(\bm{\mu\mu}^H+\bm{\Sigma})\right)}{2}+d}.
\end{equation}

Using the alternating approach, $\{\bm{\mu},\bm{\Sigma}\}$ and the hyperparamters can be iteratively optimized to convengency. Finally, $\bm{\mu}$ is regarded as the final point estimate for $\bm{\xi}$.
The main complexity is dominated by the inverse opertor for $\bm{\Sigma}$, incurring a complexity of $\mathcal{O}(G^3)$ per iteration. Note that this can be reduced by applying the Woodbury formula.
\subsubsection{MFV-SBL} 

MFV-SBL \cite{MFS1,MFS2}, also called space alternating variational estimation (SAVE)-SBL in \cite{MFS1}, considered a fully factorized form over $\bm{\Theta}$ extending  (\ref{qt}) further:
\begin{equation}
	q(\bm{\Theta})=\prod_{i=1}^{G}q(\xi_i)\prod_{i=1}^Gq(\rho_i)q(\sigma^2).
\end{equation}

This will significantly decrease the complexity by avoiding the matrix inverse when updating parameters.
For each scalar variable $\Theta_u$, $u\in\{1,\cdots, 2G+1\}$, in $\bm{\Theta}$, we obtain optimal $q(\Theta_u)$:
\begin{equation}\label{qt2}
	q(\Theta_u)=\frac{e^{\mathbb{E}_{\bm{\Theta}_{\lnot u}} \{\log p(\mathbf{y},\bm{\Theta})\}}}{\int e^{\mathbb{E}_{\bm{\Theta}_{\lnot u}} \{\log p(\mathbf{y},\bm{\Theta})\}} {\rm d}\Theta_u},
\end{equation}
where $\bm{\Theta}_{\lnot u}$ denotes all variables in $\bm{\Theta}$ except for $\Theta_u$.
 
Combining (\ref{pyt}) and (\ref{qt2}), the mean and covariance of $q(\xi_i)$ can be derived:
\begin{equation}
	\mu_i =\mathbb{E}\{\sigma^{-2}\} [\bm{\Sigma}]_{i,i} [\bm{\Phi}]_{:,i}^H (\mathbf{y}-\bm{\Phi}_{\lnot i}\bm{\mu}_{\lnot i}),
\end{equation}
\begin{equation}\label{SIG}
[\bm{\Sigma}]_{i,i}=\frac{1}{\mathbb{E}\{\sigma^{-2}\}\Vert[\bm{\Phi}]_{:,i}\Vert_2^2+\rho_i}.
\end{equation}

Comparing  (\ref{var2}), the matrix invese is avoided thanks to the fully factorized distribution and iterative updating for $[\bm{\Sigma}]_{i,i}$.
Moreover, the updating rule for parameters $\bm{\rho}$ and $\sigma^2$ are the same as (\ref{rho2}) and (\ref{s2}).

Given some pre-calculations and storage memory \cite{MFS1}, the main complexity of MFV-SBL, which is $\mathcal{O}(G^2)$ per iteration, arises primarily from $N$ matrix-vector multiplications needed to calculate $\{\mu_i\}_{i=1}^G$.

\subsubsection{Proposed CMFV-SBL} 
There are some drawbacks in V-SBL and MFV-SBL. V-SBL focuses on full-dimension information, introducing high complexity. MFV-SBL, on the other hand, updates only one variable at a time in a greedy manner, which may struggle with highly correlated cases. To address this, we propose CMFV-SBL. This method strikes a balance by using partially factorized distributions. It divides the variables $\bm{\xi}$ and $\bm{\rho}$ into $K$ clusters, such that $\bm{\xi} \rightarrow \{\bm{\xi}_1, \cdots, \bm{\xi}_K\}$ and $\bm{\rho} \rightarrow \{\bm{\rho}_1, \cdots, \bm{\rho}_K\}$.
\begin{equation}
	q(\bm{\Theta})=\prod_{k=1}^{K}q(\bm{\xi}_k)\prod_{k=1}^Kq(\bm{\rho}_k)q(\sigma^2).
\end{equation}
 
 Extending  (\ref{qt2}) into this case, we can obtain the optimal $q(\bm{\xi}_k)$:
 \begin{equation} 	q(\bm{\xi}_k)=\frac{e^{\mathbb{E}_{\bm{\xi}_{\lnot k},\bm{\rho},\sigma^2} \{\log p(\mathbf{y},\bm{\Theta})\}}}{\int e^{\mathbb{E}_{\bm{\xi}_{\lnot k},\bm{\rho},\sigma^2} \{\log p(\mathbf{y},\bm{\Theta})\}} {\rm d} \bm{\xi}_k}.
 \end{equation}
By substituting the residual $\mathbf{r} = \mathbf{y} - \bm{\Phi}_{\lnot k}\bm{\xi}_{\lnot k}$, where $\bm{\Phi}_{\lnot k}$ is formed by the columns in $\bm{\Phi}$ that do not include the columns corresponding to $\bm{\xi}_k$, into  (\ref{pyt}), the quadratic term becomes
\begin{equation}
	\frac{1}{\sigma^2}\Vert \mathbf{y}-\bm{\Phi}\bm{\xi}\Vert^2_2=	\frac{1}{\sigma^2}\Vert \mathbf{r}-\bm{\Phi}_k\bm{\xi}_k\Vert^2_2.
\end{equation}

Therefore, we can obtain the mean and covariance of $\bm{\xi}_k$, $\forall k$:
\begin{equation}
	\bm{\mu}_k=\mathbb{E}\{\sigma^{-2}\} \bm{\Xi}_k \bm{\Phi}_k^H\mathbf{r},
\end{equation}
\begin{equation}
	\bm{\Sigma}_k=\left(\mathbb{E}\{\sigma^{-2}\}\bm{\Phi}_{k}^H\bm{\Phi}_k+{\rm diag}(\bm{\rho}_k)
	\}
	\right)^{-1}.
\end{equation}

Compared to the scalar inverse in  (\ref{var2}), only a low-dimensional matrix inverse is required here. To analyze the complexity, we assume $D$ clusters are uniformly divided. The complexity of calculating $\bm{\mu}$ and $\bm{\Sigma}$ per iteration is $\mathcal{O}(\max(G(G-D), GD^2))$. Here, $D=G$ and $D=1$ correspond to V-SBL and MFV-SBL, respectively. The updating rules for the parameters $\bm{\rho}$ and $\sigma^2$ are the same as in  (\ref{rho2}) and (\ref{s2}).
The selection of the number of clusters and the elements clustered together can significantly impact the performance of CMFV-SBL. We apply the K-means clustering approach to divide the $G$ atoms into $D$ clusters. Additionally, to accelerate the algorithm, we remove indices with very large $\rho$ values in each iteration, thereby reducing the dictionary size as the iterations progress.

\begin{figure}
	\centering
	\subfigure[Cascaded Channel.]{
		\includegraphics[width=2.8in]{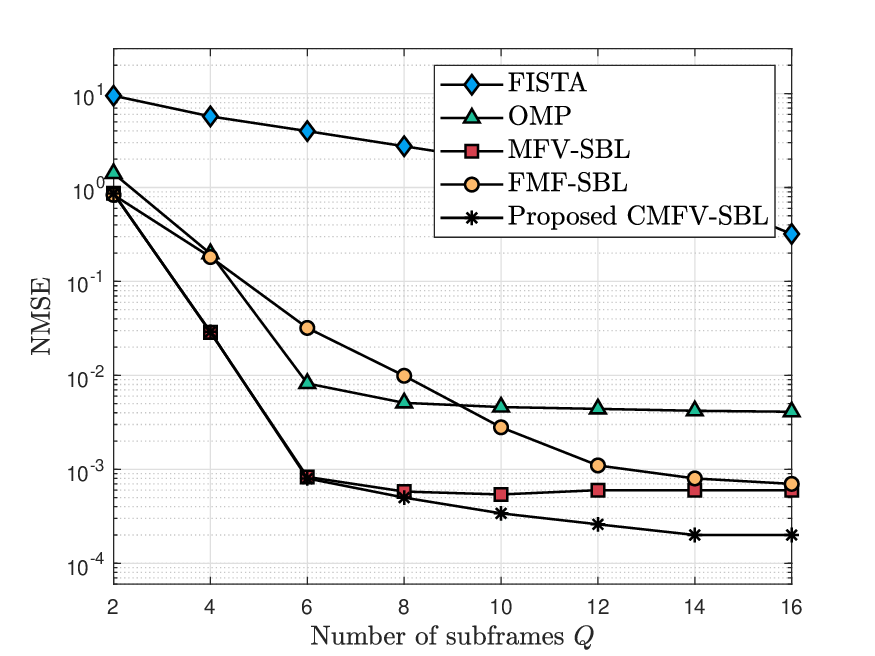}
	}
	\\    
	\subfigure[Direct Channel.]{
		\includegraphics[width=2.8in]{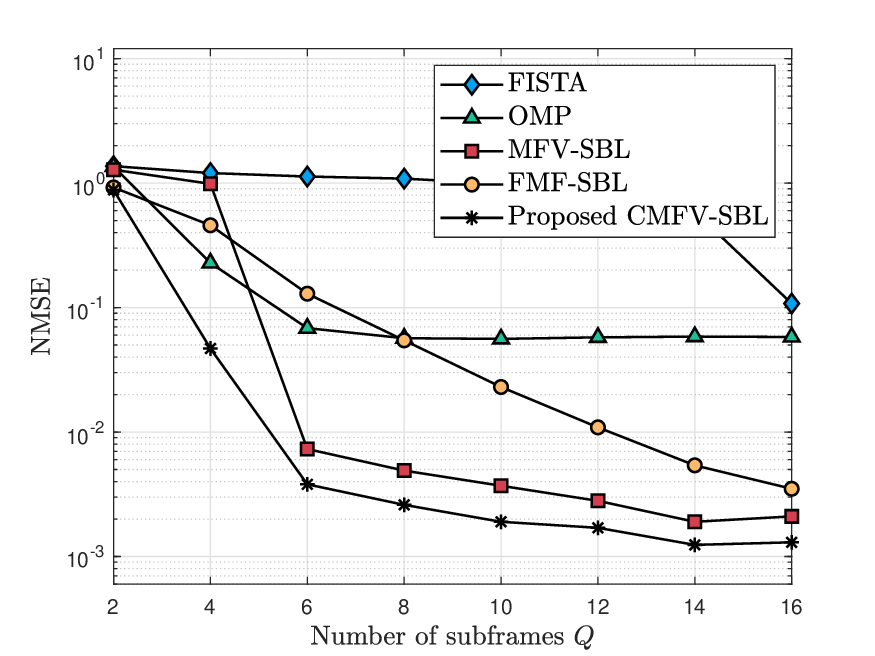}
	} 
	\caption{NMSE of different methods versus $Q$.}
	\label{nmse_q}
\end{figure}
\section{Simulation Results} 
In this section, we conduct a series of numerical simulations to evaluate channel estimation for FIM. The simulations are set in a system operating at a central frequency of $10$ GHz. The FIM is equipped with $N =4 \times 4$ movable elements. Path gains $\alpha_l,\beta_{p}$,   $\forall l,p$, are modeled following a complex Gaussian distribution, $\mathcal{CN}(0,1)$.  The number of channel paths $L$ and $P$ are set to 4. The azimuth angle $\widetilde{\phi}_{l,p}$ and elevation angle $\widetilde{\theta}_{l,p}$ are assumed to distribute on a $12\times 12$ grid. The time slots of EM are set to $T_2=9$, which make the $4\times 4$ elements move to form a $12\times 12$ virtual array for channel estimation. 
 The signal-to-noise ratio (SNR) is defined as $1/\sigma^2$. In the following simulations, SNR and the subframe $Q$ are varied to evaluate the normalized mean squared error (NMSE) performance for cascaded and direct channel estimation, where cascaded and direct channels are characterized by the virtual array size and a scalar, respectively.
This paper evaluates various methods, including fast iterative shrinkage-thresholding algorithm (FISTA) \cite{FISTA}, orthogonal matching pursuit (OMP) \cite{OMP}, MFV-SBL \cite{MFS1}, FMF-SBL \cite{MFS2}, and the proposed CMFV-SBL. For all algorithms except OMP, the maximum number of iterations is set to 400. The iteration count for OMP is set to match the number of cascaded paths, $PL=16$.
 
 \begin{figure}
 	\centering
 	\subfigure[Cascaded Channel.]{
 		\includegraphics[width=2.8in]{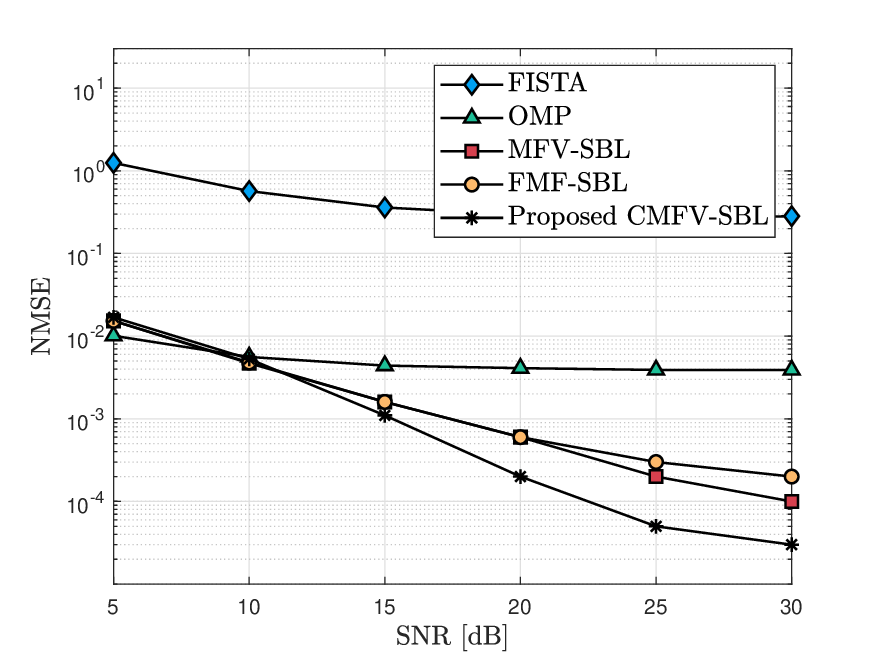}
 	}
 	\\    
 	\subfigure[Direct Channel.]{
 		\includegraphics[width=2.8in]{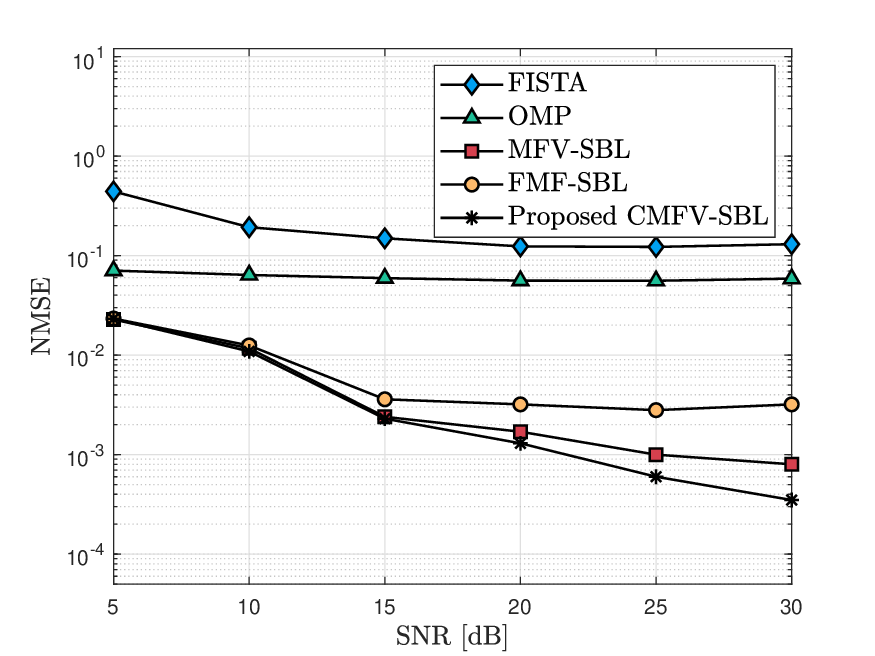}
 	} 
 	\caption{NMSE of different methods versus SNR.}
 	\label{nmse_snr}
 \end{figure}
 We first evaluate the impact of the number of subframes $Q$ on different methods. As shown in Fig. \ref{nmse_q}, where $Q$ ranges from 2 to 16 and the SNR is set to 20 dB, all algorithms except FISTA exhibit substantial NMSE performance for both cascaded and direct channel estimation, highlighting the effectiveness of the proposed joint estimation framework. The NMSE trend indicates that within the proposed framework, low pilot overhead is sufficient for channel estimation, with $Q = 8$  achieving significant NMSE performance. Furthermore, the simulations demonstrate that the proposed CMFV-SBL algorithm outperforms other benchmarks in joint cascaded and direct channel estimation.

 We then evaluate the impact of SNR values on different methods. As shown in Fig. \ref{nmse_snr}, with SNR ranging from $5$ to $30$ dB and the number of subframes being set to $Q = 16$, it can be observed that FISTA and OMP show slight NMSE performance improvements for both cascaded and direct channel estimation as SNR increases. At low SNR levels, below $15$ dB, FMF-SBL, MFV-SBL, and CMFV-SBL exhibit comparable performance. However, as SNR increases, CMFV-SBL demonstrates a clear advantage over the other two algorithms.

\begin{figure}
   	\centering 
   	\includegraphics[width=3in]{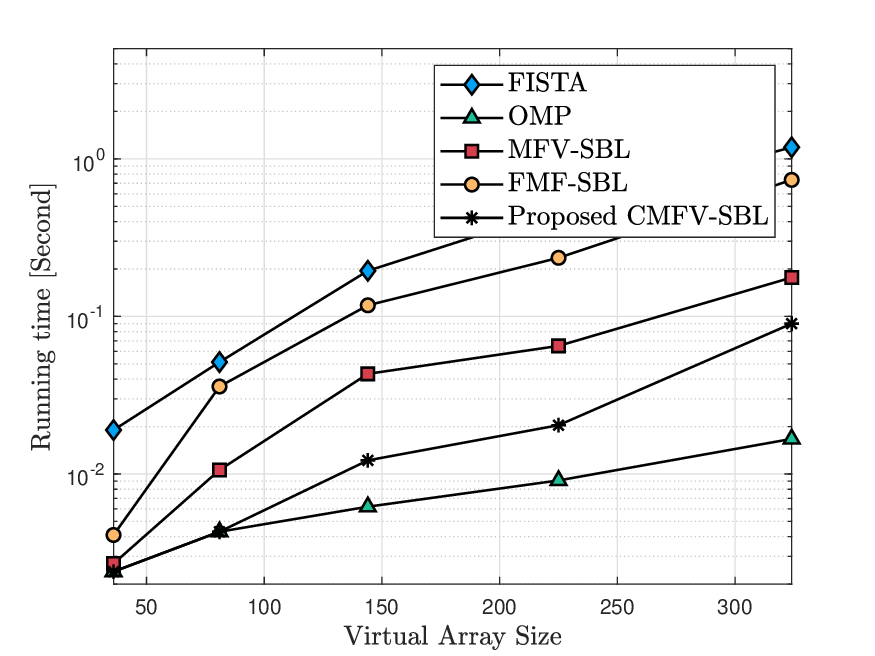}
   	\caption{Running time versus the virtual array size.}\label{TIME} 
   \end{figure}

  Fig. \ref{TIME} presents the running time for different methods using a 13th Gen Intel(R) Core(TM) i7-13650HX CPU to illustrate their time complexity. The number of subframes $Q$ and the SNR are set to $12$ and $20$ dB, respectively. The virtual array is formed by multiple EMs with a size of $NT_2$. The maximum iteration number for FISTA, MFV-SBL, FMF-SBL, and the proposed CMFV-SBL is set to $400$, with the tolerance for stopping error set to $10^{-8}$. It can be observed that the OMP algorithm has the fastest speed, as it typically requires fewer iterations in highly sparse cases. However, its drawbacks include reliance on known sparsity and relatively poor performance. Notably, the proposed CMFV-SBL algorithm achieves the second fastest speed and is comparable to OMP when the virtual array size is small.
   
\section{Conclusions}\label{Con}
This paper investigates FIM-aided communications in a SISO setup. First, the EM-only, PBF-only, and EM-PBF modes are compared in terms of received signal power, demonstrating that: 1) spatial constructive and destructive interference effects depend on element positions and phase coefficients, 2) the EM-only mode, which optimizes element positions, outperforms the PBF-only mode, which optimizes phase coefficients, by effectively addressing multi-path effects, and 3) the EM-PBF mode achieves superior performance. Additionally, a channel estimation protocol for single- and multi-element FIM is designed, along with a joint cascaded and direct channel estimation framework within a sparse recovery problem. To this end, we propose a clustering mean-field variational sparse Bayesian learning algorithm, which proves the effectiveness of the estimation protocol and framework, and outperforms the benchmarks.

\begin{appendices}
	\section{Proof of Proposition \ref{ta1}}\label{A1}
	We first simplify the expressions of $f_{\rm PBF}(v^\star)$ and $	f_{\rm UB}$ as: 
	\begin{equation}
		\begin{aligned}
			f_{\rm PBF}(v^\star)=&\left(\vert\gamma\vert+\frac{1}{\sqrt{LP}} \left\vert  \sum_{l=1}^{L}\sum_{p=1}^{P} \alpha_l\beta_p^\star
			\right\vert
			\right)^2 \\
			=& \vert\gamma\vert^2+ \frac{2}{\sqrt{LP}}\vert\gamma\vert\left\vert  \sum_{l=1}^{L}\sum_{p=1}^{P} \alpha_l\beta_p^\star
			\right\vert + \frac{1}{LP} \left\vert  \sum_{l=1}^{L}\sum_{p=1}^{P} \alpha_l\beta_p^\star
			\right\vert^2,
		\end{aligned}
	\end{equation} 
	\begin{equation}
		\begin{aligned}
			f_{\rm UB} &=\left(|\gamma|+ \frac{1}{\sqrt{LP}} \sum_{l=1}^{L}\sum_{p=1}^{P}|\alpha_l\beta^*_p| \right)^2 \\
			& = |\gamma|^2+ \frac{2}{\sqrt{LP}} |\gamma|\sum_{l=1}^{L}\sum_{p=1}^{P} |\alpha_l\beta^*_p| + \frac{1}{LP}\left(\sum_{l=1}^{L}\sum_{p=1}^{P}|\alpha_l\beta^*_p|\right)^2.
		\end{aligned}
	\end{equation}
	
	We assume $\gamma\sim \mathcal{CN}(0,\sigma^2_\gamma)$,
	$\alpha_l\sim \mathcal{CN}(0,\sigma^2_\alpha)$, and
	$\beta_p\sim \mathcal{CN}(0,\sigma^2_\beta)$, such that $\sum_{l=1}^{L}\sum_{p=1}^{P} \alpha_l\beta_p^\star \sim \mathcal{CN}(0,LP\sigma_\alpha^2\sigma_\beta^2)$, the amplitude $\vert\gamma\vert \sim \text{Rayleigh}\left(\frac{\sigma_\gamma}{\sqrt{2}}\right)$, $\left\vert  \sum_{l=1}^{L} \alpha_l
	\right\vert\sim \text{Rayleigh}\left( \frac{\sqrt{L}\sigma_\alpha}{\sqrt{2}}\right)$, and  $\left\vert  \sum_{p=1}^{P} \beta^*_p
	\right\vert\sim \text{Rayleigh}\left( \frac{\sqrt{P}\sigma_\beta}{\sqrt{2}}\right)$, where   $ \text{Rayleigh}\left(\sigma\right)$ is the Rayleigh distribution with scale parameter $\sigma$.

	The expectation of  $f_{\rm PBF}(v^\star)$ is given by
	\begin{equation}
		\begin{aligned} 
			\mathbb{E}\left\{ f_{\rm PBF}(v^\star)
			\right\}=& \mathbb{E}\left\{\vert\gamma\vert^2	\right\}+\frac{2}{\sqrt{LP}}\mathbb{E}\left\{\vert\gamma\vert\left\vert  \sum_{l=1}^{L}\sum_{p=1}^{P} \alpha_l\beta_p^\star
			\right\vert	\right\}
			\\ &+ \frac{1}{LP}\mathbb{E}\left\{ \left\vert  \sum_{l=1}^{L}\sum_{p=1}^{P} \alpha_l\beta_p^\star
			\right\vert^2
			\right\} \\
			=&
			\sigma_\gamma^2+ \frac{2}{\sqrt{LP}}\mathbb{E}\left\{ \vert\gamma\vert	\right\}\mathbb{E}\left\{ \left\vert\sum_{l=1}^{L}\alpha_l\right\vert
			\right\}\mathbb{E}\left\{ \left\vert\sum_{p=1}^{P}\beta_p\right\vert
			\right\} \\
			&+\sigma_\alpha^2\sigma_\beta^2 \\
			\overset{(c)}{=}&	\sigma_\gamma^2+\frac{\pi\sqrt{\pi}}{4}\sigma_\gamma\sigma_\alpha\sigma_\beta
			+	\sigma_\alpha^2\sigma_\beta^2,
		\end{aligned}
	\end{equation}
	where $(c)$ holds due to $\mathbb{E}\left\{  \text{Rayleigh}\left(\sigma\right)
	\right\}=\sqrt{\frac{\pi}{2}}\sigma$.
	
The expectation of  $f_{\rm UB}$ is given by
	\begin{equation}
		\begin{aligned}
			\mathbb{E}\left\{ f_{\rm UB}
			\right\} =& 
			\mathbb{E}\left\{|\gamma|^2\right\}+ \frac{2}{\sqrt{LP}} \mathbb{E}\left\{ |\gamma|\sum_{l=1}^{L}\sum_{p=1}^{P} |\alpha_l\beta^*_p|\right\}  \\&+ \frac{1}{LP}\mathbb{E}\left\{\left(\sum_{l=1}^{L}\sum_{p=1}^{P}|\alpha_l\beta^*_p|\right)^2\right\} \\
			=& \sigma_\gamma^2 +\frac{2}{\sqrt{LP}}\mathbb{E}\left\{ \vert\gamma\vert	\right\}\mathbb{E}\left\{ \sum_{l=1}^{L}\left\vert\alpha_l\right\vert
			\right\}\mathbb{E}\left\{ \sum_{p=1}^{P}\left\vert\beta_p\right\vert
			\right\} \\ 
			&+ \frac{1}{LP}\mathbb{E}\left\{ \left( \sum_{l=1}^{L}\vert \alpha_l\vert\right)^2
			\right\}\mathbb{E}\left\{ \left( \sum_{p=1}^{P}\vert \beta_p\vert\right)^2
			\right\}.
		\end{aligned}
	\end{equation}
	
	Furthermore, we have
	\begin{equation}
		\begin{aligned}
			\mathbb{E}\left\{ \left( \sum_{l=1}^{L}\vert \alpha_l\vert\right)^2
			\right\}= &\mathbb{E}\left\{ \sum_{l=1}^{L} \vert\alpha_l\vert^2 +2\sum_{l_1<l_2}^{L}\vert \alpha_{l_1}\vert \vert\alpha_{l_2}\vert 
			\right\} \\
			\overset{(d)}{=}& L\sigma^2_\alpha + L(L-1) \frac{ \pi\sigma^2_\alpha}{4},
		\end{aligned}
	\end{equation}
		\begin{equation}
		\begin{aligned}
			\mathbb{E}\left\{ \left( \sum_{l=1}^{L}\vert \alpha_l\vert\right)^2
			\right\} 
			\overset{(d)}{=}  P\sigma^2_\beta + P(P-1) \frac{ \pi\sigma^2_\beta}{4},
		\end{aligned}
	\end{equation}
	where $(d)$ holds due to the known variance of complex Gaussian and the expectation of Rayleigh distribution.
	
{Hence, we obtain}
	\begin{equation}
		\begin{aligned}
			\mathbb{E}\left\{f_{\rm UB}\right\}=&\sigma_\gamma^2 + \frac{\pi\sqrt{\pi}\sqrt{LP}}{4} \sigma_\gamma\sigma_\alpha\sigma_\beta
			\\
			& +\left(1+\frac{ \pi (L-1) }{4}\right)\left(1+\frac{ \pi (P-1) }{4}\right)\sigma^2_\alpha\sigma^2_\beta.
		\end{aligned}
	\end{equation}
\end{appendices}
\bibliographystyle{IEEEtran}
\bibliography{reference.bib}

\vspace{12pt}

\end{document}